\documentclass[twocolumn]{aastex631}

\begin{document}

\title{Danger Zone: Establishing Buffers for Enhanced Classification in BPT Diagrams}

\author[0000-0002-9879-1749]{Changhyun Cho}
\affiliation{New York University Abu Dhabi, PO Box 129188, Abu Dhabi, UAE}
\affiliation{Center for Astrophysics and Space Science (CASS), New York University Abu Dhabi, PO Box 129188, Abu Dhabi, UAE}
\affiliation{New York University, 726 Broadway, New York, NY 10003}
\affiliation{Department of Physics and Astronomy, York University, 4700 Keele Street, Toronto, ON M3J 1P3, Canada}

\author[0000-0002-9220-0039]{Ahmad Nemer}
\affiliation{New York University Abu Dhabi, PO Box 129188, Abu Dhabi, UAE}
\affiliation{Center for Astrophysics and Space Science (CASS), New York University Abu Dhabi, PO Box 129188, Abu Dhabi, UAE}

\author[0000-0002-6425-6879]{Ivan Yu. Katkov}
\affiliation{New York University Abu Dhabi, PO Box 129188, Abu Dhabi, UAE}
\affiliation{Center for Astrophysics and Space Science (CASS), New York University Abu Dhabi, PO Box 129188, Abu Dhabi, UAE}
\affiliation{Sternberg Astronomical Institute, Lomonosov Moscow State University, Universitetskij pr., 13,  Moscow, 119234, Russia}

\author[0000-0003-4679-1058]{Joseph D. Gelfand}
\affiliation{New York University Abu Dhabi, PO Box 129188, Abu Dhabi, UAE}
\affiliation{Center for Astrophysics and Space Science (CASS), New York University Abu Dhabi, PO Box 129188, Abu Dhabi, UAE}
\affiliation{Center for Cosmology and Particle Physics, New York University, 726 Broadway, Room 958, New York, NY 10003}

\begin{abstract}
This study utilizes unsupervised machine learning, specifically the uniform manifold approximation and projection (UMAP) algorithm, to classify optical spectra originating from star-forming regions, Seyferts, and low-ionization (nuclear) emission-line regions (LI(N)ERs) based on their line ratios. Typically, the ionization source of a region is determined from intensity ratio of different combinations of pairs of spectral lines. However, using current boundary definitions, $\sim10$\% of spectra change classes between diagnostic diagrams. We apply the machine learning technique to $\sim$1.3 million optical spectra from 6 439 galaxies observed in the MaNGA survey. By training UMAP on consistently classified data, we can classify these ``ambiguous'' spectra, and delineate boundary zones where such ambiguities arise. Furthermore, we identify physically interesting subsets within the ambiguous spectra. Future work will incorporate additional parameters, such as alternative emission line ratios and velocity dispersions, to enhance classification accuracy.%

\end{abstract}

\keywords{galaxy evolution, spectroscopy, line emission, machine learning}

\section{Introduction} \label{sec:intro}
BPT diagrams, introduced by Baldwin, Phillips, and Terlevich, are fundamental tools in astrophysics for classifying the optical emission-line spectra of ionization regions within galaxies \citep{baldwi1981, veille1987}. By comparing specific line ratios, these diagrams help distinguish between ionization sources such as star-formation, active galactic nuclei (AGN) activity, and shocks. Because emission-line ratios are sensitive to parameters such as the ionization level and the hardness of the ionizing radiation field, BPT diagrams provide valuable insight into the energetic processes that shape galaxy evolution \citep[e.g.,][]{kewley2013a, kewley2019}.

The observed emission lines originate from hot ($T \gtrsim 8 \ 000$ K) ionized gas (e.g., H$\alpha$, H$\beta$, [N II], [O II] and [O III]) or neutral gas (e.g., [O I]), provide useful information into the physical conditions and nature of the ionization sources within galaxies \citep{osterbrock2006, kewley2006}. These lines reflect the spectrum of incident photons that heat or photo-ionize the gas, revealing the presence of underlying star-formation, AGN activity, shocks, and/or post-AGB stars \citep{cidfernandes2011, rich2011, kewley2013b}. Emission line ratios are especially useful because they reduce the impact of dust reddening; since lines of similar wavelengths experience similar reddening, their ratios help cancel out much of the dust’s effect. This differentiation helps classify regions into star-forming regions (SF), AGN, Seyferts (Sy), or low-ionization (nuclear) emission-line regions (LI(N)ERs) \citep{baldwi1981, kewley2001}.

BPT diagrams enhance the classification of ionization regions by employing demarcation curves \citep{kewley2001, kauffm2003, kewley2006, RN242}. The most widely used division lines, proposed by \cite{kewley2001}, \cite{kewley2006} and \cite{kauffm2003}, effectively distinguish between SF, Sy, LI(N)ERs, and composite zones. \cite{kewley2001} introduced a boundary between SF and AGN, derived from stellar population synthesis and photoionization models, representing the theoretical maximum starburst line. \cite{kauffm2003} further added an empirical line separating pure SF from Seyfert--H II composites. Subsequently, \cite{kewley2006} refined these boundaries by fitting through histogram minima in the diagnostic diagrams, providing a clearer empirical separation between Sy and LI(N)ERs.

While the classification effectively differentiates SF from AGN, it encounters ambiguity in certain parts of the BPT diagram. Specifically, the boundaries separating Sy from LI(N)ERs and SF from composites are based on empirical definitions rather than physical models \citep{kauffm2003, kewley2006}. As a result, many regions exhibit spectra that fall near these boundaries, making it challenging to determine their true classification. For instance, \citet{maksym2016} found that most regions within the inner kiloparsec of NGC 3393 have spectra straddling the boundary between Sy and LI(N)ER regions. This highlights the need for a more refined classification method to better distinguish these ionization regions.

The ambiguity in identifying the origin of incident photons stems from a mixture of photons from various sources, leading to inconsistent classifications across BPT diagrams. For instance, a spaxel classified as SF in one diagram might be labeled as Sy or LI(N)ER in another. These inconsistencies may also reflect limitations in the demarcation lines themselves, some of which have been revised in recent studies to better match observational trends \citep[e.g.,][]{law2021}. The issue is particularly evident in Seyfert galaxies, where optical spectra often feature emission lines excited by both star-formation and AGN activity, complicating interpretation. Although H$\alpha$ luminosity in H II regions is typically proportional to the star-formation rate \citep[e.g.,][]{kennicutt1998}, it can also be influenced by AGN ionizing radiation. Additionally, \citet{stasin2006} reported that BPT diagrams may be less effective in detecting AGN within metal-poor galaxies, as the \citet{kauffm2003} line includes SF regions that could have an AGN contribution to H$\beta$. Similarly, \citet{trump2015} showed that intense star-formation can dilute AGN signatures on BPT diagrams, further complicating the classification in composite systems.

Efforts have been made to overcome the limitations of BPT diagrams and enhance our understanding of ionization regions more effectively. The advent of integral field spectroscopy (IFS) in recent large-scale surveys, such as CALIFA, MaNGA, and SAMI \citep{califa2012, manga2015, sami2021}, has made it possible to acquire highly detailed spatial data. This advancement allows for closer examination of ionization regions within galaxies, revealing that a single galaxy can display a range of ionization types—SF, Sy, and LI(N)ER—depending on local environmental conditions. IFS enables the study of thousands of nearby galaxies with rich, multidimensional datasets at sub-galactic scales. Notably, recent studies such as \citet{alban2023} have leveraged MaNGA’s spatially resolved data to identify AGN more robustly within galaxies, further demonstrating the power of IFS in disentangling complex ionization structures.

At the same time, the need for efficient management of multidimensional and large data volumes has grown increasingly important. Integrating advanced data processing techniques, including machine learning, with IFS enables more robust and efficient analysis of ionization structures, leading to a more refined interpretation of emission line properties within galaxies. Machine learning algorithms, in particular, can streamline data handling and pattern recognition, offering powerful tools for extracting complex insights from vast datasets.

Several previous studies have aimed to refine emission-line diagnostic boundaries and introduce alternative methods for classifying gas excitation sources in galaxies. For example, \citet{law2021_refine} used velocity dispersions and additional properties to revise traditional BPT diagram boundaries. They demonstrated that their demarcation curves effectively distinguish different ionization sources, both in traditional two-dimensional (2D) diagrams and in expanded three-dimensional (3D) representations. Other studies have explored the potential usefulness of 3D line-ratio diagnostic diagrams. For instance, \citet{vogt2014} introduced $\mathcal{ZQE}$ diagrams, which separate the oxygen abundance and the ionization parameter of H II region-like spectra, while also probing the excitation mechanisms of the gas. \citet{dagostino2019} and \citet{RN170} also developed 3D diagnostic diagrams that incorporate kinematic information such as velocity dispersion to differentiate between photoionization and shock excitation.

In addition to emission-line ratios, recent studies have explored dimensionality reduction methods using full galaxy spectra to classify gas excitation regions. Approaches such as variational autoencoders \citep[VAEs;][]{portillo2020}, probabilistic auto-encoders \citep[PAE;][]{boehm2022probabilistic, pat2020}, and SPENDER \citep{melchior2023} leverage complete spectral data rather than limited emission-line ratios. These methods reconstruct spectral information and aim to represent BPT classifications and sample distributions within simplified 2D spaces. However, spectral-based techniques are more sensitive to noise, typically requiring larger sample sizes, dust-absorption corrected data, and complex architectures that can be difficult to modify.

Alternatively, some studies have employed dimensionality reduction techniques using selected emission-line ratios instead of full spectra. Using emission-line ratios offers advantages, such as reduced sensitivity to dust absorption effects, thereby requiring comparatively smaller sample sizes. Additionally, because significant emission lines used in traditional diagnostics like the BPT diagrams are already selected, these methods do not require the encoding process inherent in spectral-based methods. \citet{zhang} demonstrated that the t-distributed Stochastic Neighbor Embedding \citep[t-SNE;][]{tsne} effectively differentiates between type-2 AGN and star-forming galaxies. However, while t-SNE successfully delineates between these broad categories, it struggles to identify finer substructures such as distinguishing Seyfert galaxies from LI(N)ERs. Limitations of t-SNE include its emphasis on local data structures at the expense of global relationships, making it less effective for interpreting complex dataset structures comprehensively. Furthermore, the computational intensity of t-SNE restricts its scalability for extensive datasets \citep{tsne}.

Uniform Manifold Approximation and Projection (UMAP) is a non-linear dimensionality reduction technique based on manifold learning and graph theory \citep{umap}. It offers several advantages over t-SNE, including faster computation, improved preservation of global structures, and better scalability with large datasets. UMAP’s flexibility enables it to handle various data types and distance metrics, making it suitable for both unsupervised and supervised learning tasks. It has been successfully applied in different astrophysical classifications, such as galaxy morphologies \citep{RN235}, Type II supernovae \citep{RN236}, gamma-ray bursts \citep{RN237}, and fast radio bursts \citep{RN199}.

In this study, we utilized the UMAP algorithm to analyze our dataset, focusing on a four-dimensional parameter space composed of the widely-known BPT ratios; [O III]/H$\beta$, [N II]/H$\alpha$, [S II]/H$\alpha$, and [O I]/H$\alpha$. Testing the UMAP-trained model revealed distinct clusters similar to those in traditional BPT diagrams, while also suggesting that this method could be valuable for identifying individual cases of interest. Section \ref{sec:manga} outlines the MaNGA data collection and sample selection process employed for the machine learning analysis. In Section \ref{sec:model}, we explain the implementation of the UMAP algorithm, including its training, testing with MaNGA samples, and subsequent clustering methods. Section \ref{sec:results} presents a comparative analysis of our model’s results against classical emission-line diagnostics, showing the advantages of the UMAP model. We also discuss further usefulness of this approach by adding additional parameters for the classification in Section \ref{sec:disc}. Finally, Section \ref{sec:conclusion} concludes our findings, their implications, and provides future plan.

\section{Data Samples}
\label{sec:manga}

\subsection{MaNGA Integral Field Spectroscopy}

The Mapping Nearby Galaxies at Apache Point Observatory (MaNGA) survey \citep{manga2015} comprises of integral field unit (IFU) \citep{ifu2015} observations of $\sim$ 10 000 galaxy with the Sloan 2.5 m optical telescope \citep{gunn2006}. Each galaxy was observed using buffered fibers with 120-micron (2$^{\prime\prime}$) core diameters. The sizes of the IFUs range from 19 to 127 fibers, with diameters varying from 12$^{\prime\prime}$ to 32$^{\prime\prime}$, matched to observed size of galaxy. Fiber bundles feed two dual-channel BOSS spectrographs \citep{smee2013}, covering a wavelength of $\lambda \lambda$ 3 600—13 000 $\text{\AA}$ with a resolving power $R \sim 2 000$, which corresponds to an instrumental dispersion of $\sigma_{\text{inst}} \approx 75 \text{ km s}^{-1}$ at 5 100 $\text{\AA}$ \citep{law2016}. To ensure complete coverage, each galaxy was observed using three dithered exposures,  effectively filling the gaps between the fibers within each bundle \citep{law2015, yan2016}.

In this work, we used data products available from the Data Analysis Pipeline (DAP) of the MaNGA survey. The DAP provides a comprehensive set of properties extracted from the MaNGA spectral cubes, including stellar and ionized gas kinematics, emission line fluxes, and other parameters \citep{westfa2019,belfio2019,law2021}. Specifically, we utilized the emission line measurements of several strong lines obtained using Gaussian parameterization. We used these values from DAP (version 3.1.0) maps computed using the \textsc{HYB10-MILESHC-MASTARSSP} binning schema, which achieves a signal-to-noise of 10 in the continuum during full spectral fitting while retaining spatial resolution for emission lines to capture ionized gas properties accurately.

\subsection{Sample selection (Training and testing sets)}
\label{sec:sample}

\subsubsection{Training set (Clean data)}
\label{subsec:clean}

\begin{figure*}
\plotone{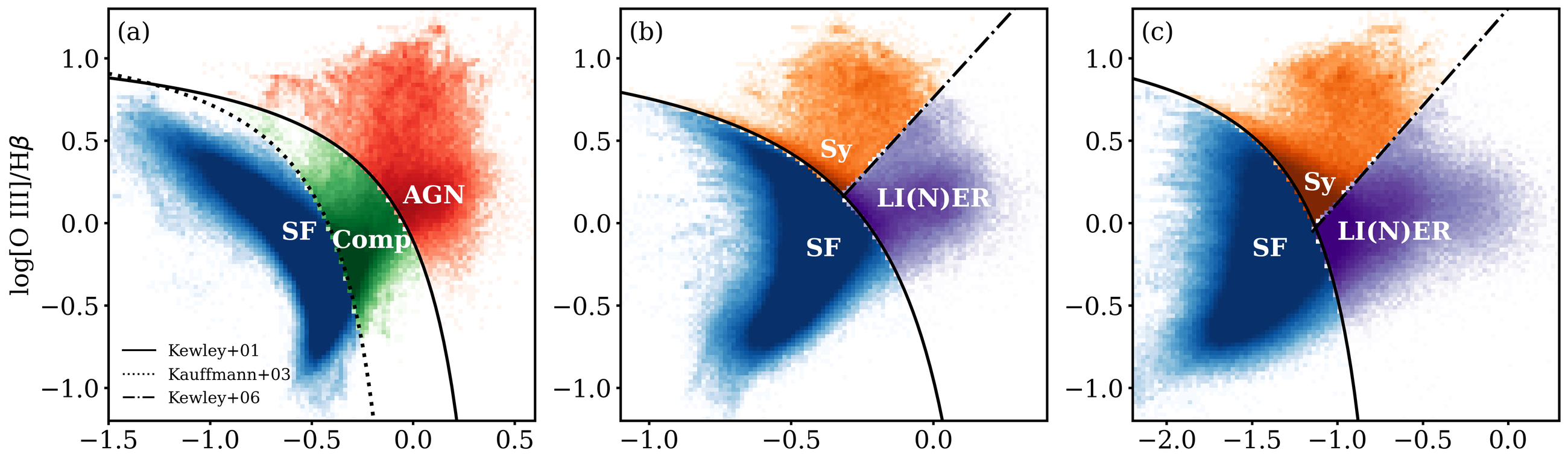}
\plotone{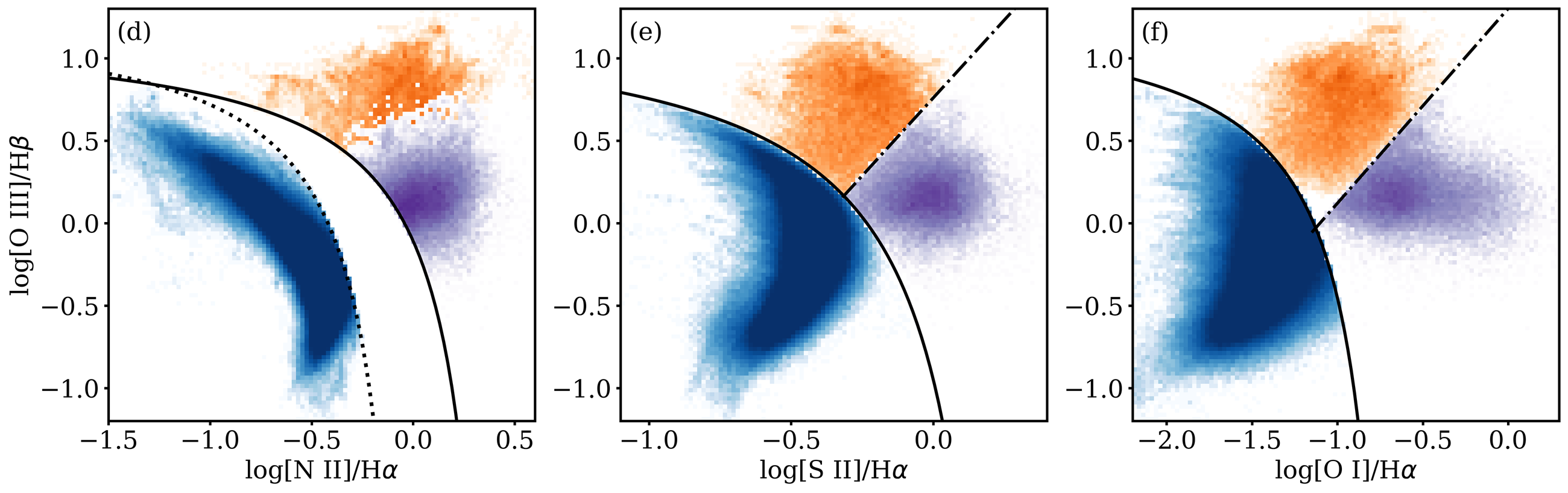}
\caption{Three BPT diagnostic diagrams for MaNGA spaxels. The upper panels (a–c) include all spaxels, whereas the lower panels (d–f) display only those spaxels consistently classified into the same category across all three diagrams, which constitute the training set for machine learning analysis. Demarcation curves defining different ionization classes are shown from \citet[][solid line]{kewley2001}, \citet[][dotted line]{kauffm2003}, and \citet[][dot-dashed line]{kewley2006}. Colors indicate classification: SF (blue), Comp (green), AGN (red), Sy (orange), and LI(N)ER (purple).}
\label{fig:bpts}
\end{figure*}

\begin{figure*}
\plotone{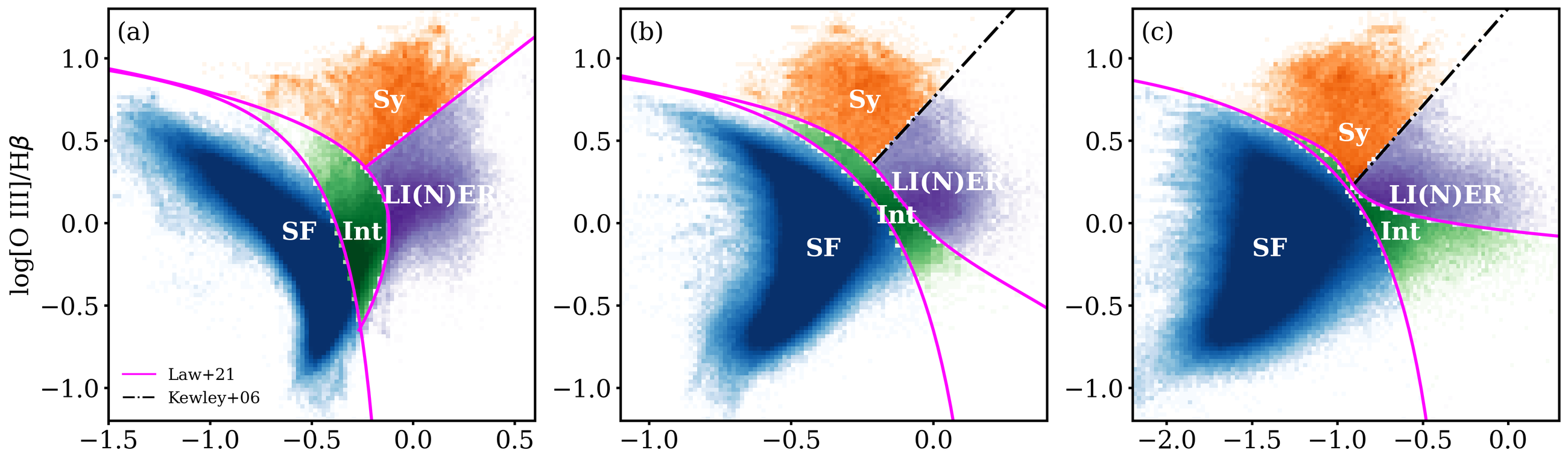}
\plotone{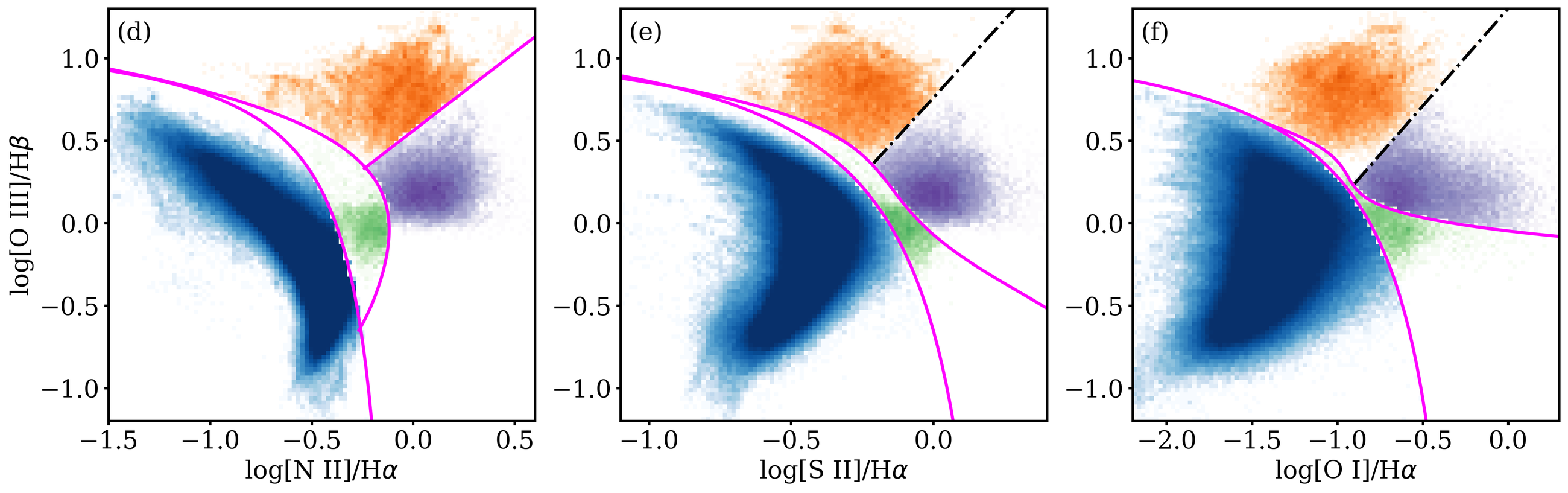}
\caption{Three BPT diagnostic diagrams for MaNGA spaxels, similar to Figure \ref{fig:bpts}, but utilizing alternative demarcation curves. The magenta curves indicate boundaries defined by \citet{law2021_refine}, while the black dot-dashed lines represent the boundary from \citet{kewley2006}. Upper panels (a–c) show all spaxels, whereas lower panels (d–f) show only the consistently classified spaxels across all three diagrams. Colors indicate classification: SF (blue), Int (green), Sy (orange), and LI(N)ER (purple).}
\label{fig:bpts_law}
\end{figure*}

For our study, we utilized the line integrated flux of $\text{H}\alpha$, $\text{H}\beta$, $[\text{N II}]\lambda6585$, $[\text{S II}]\lambda\lambda6718,32$, $[\text{O I}]\lambda6302$, and $[\text{O III}]\lambda5008$ measured by the DAP \citep{westfa2019, belfio2019, law2021}. To ensure the reliability and accuracy of our findings, we adopted a minimum signal-to-noise ratio (S/N) of 5 in the strong emission-lines used in the BPT diagrams.  While a S/N threshold may cause potential selection biases---particularly that weaker emission lines such as [O I] could preferentially exclude metal-rich SF regions---we find no significant evidence for such bias. Specifically, our BPT diagrams (Figures \ref{fig:bpts} and \ref{fig:bpts_law}) show a continuous distribution of spaxels covering the entire region associated with SF regions. According to theoretical curves by \citet{kewley2001, kewley2006}, this complete coverage indicates inclusion across the full range of metallicities. If our criteria had systematically excluded high-metallicity regions, we would observe noticeable gaps or reduced densities in the corresponding regions of the BPT diagrams, which is not the case. After applying the S/N cut, our sample encompasses data on 6 439 galaxies with a total of 1 286 655 spaxels. Among these spaxels, we define clean data as those classified unanimously as SF, Sy, or LI(N)ER across all three BPT diagrams; [O III]/H$\beta$ against [N II]/H$\alpha$, [S II]/H$\alpha$, and [O I]/H$\alpha$. This clean data set will be utilized as the training set for our unsupervised machine-learning model.

Table \ref{tab:bpts} summarizes the numbers and percentages of spaxels classified as SF, composite (Comp), Sy, and LI(N)ER across the three BPT diagrams, based on the classification boundaries defined by \citet{kewley2001}, \citet{kewley2006}, and \citet{kauffm2003}. In the [N II]-BPT diagram, approximately 85\% of spaxels are classified as SF, 6\% as AGN, and 9\% as Comp. It is important to emphasize, however, that spaxels categorized in one way in the [N II]-BPT diagram do not necessarily maintain the same classification in other BPT diagrams. Specifically, about 85\% of SF-classified spaxels from the [N II]-BPT diagram remain consistently classified in the [O I]-BPT diagram, highlighting discrepancies across classification boundaries that could significantly impact the overall classification reliability.

The six BPT diagrams in Figure \ref{fig:bpts} show the classification of spaxels into different regions: SF (blue dots), AGN (red dots), Sy (orange dots), and LI(N)ER (purple dots). In Figure \ref{fig:bpts} (a), the solid curve from \cite{kewley2001} and the dotted curve from \cite{kauffm2003} define the separation between SF, AGN, and Comp regions. The dot-dash lines in Figures \ref{fig:bpts} (b) and (c) from \cite{kewley2006} distinguish between SF, Sy, and LI(N)ERs. The upper panels depict a total of 1 286 643 spaxels with a S/N greater than 5. Notably, Comp spaxels (green dots)---identified based on the demarcation criteria of \citet{kewley2001} and \citet{kauffm2003} as regions exhibiting signatures of both star-formation and AGN activity---will be briefly addressed in Section \ref{sec:comp}.

The lower panels (d-f) in Figure \ref{fig:bpts} display spaxels consistently classified into one category across all BPT diagrams, totaling 994 304 spaxels. Note that the majority of spaxels near the classification boundaries are excluded from this consistently classified sample. This further underscores that the BPT diagrams are not fully consistent in categorizing spaxels located near the boundaries into a single class. The exclusion of these spaxels highlights the uncertainties in classification along the boundary regions, where different BPT diagrams may produce varying classifications for the same spaxels. Table \ref{tab:pure} summarizes the number of these ‘pure’ spaxels, as shown in the lower panel of Figure \ref{fig:bpts}, along with the counts of ambiguous and composite spaxels.

The inconsistency in classification is not solely related to the specific demarcation curves previously discussed; variations in demarcation methods can also significantly alter both the classification outcomes and the number of resulting categories. To further illustrate this issue, we adopt an alternative classification scheme from the literature. \citet{law2021_refine} introduced new demarcation curves based on the gas-phase velocity dispersion ($\sigma_{H\alpha}$), utilizing MaNGA spaxels to refine diagnostic boundaries. Table \ref{tab:bpts_law} summarizes the numbers and percentages of spaxels classified as SF, Intermediate (Int), Sy, and LI(N)ER according to the criteria established by \citet{law2021_refine}, across the three BPT diagrams. Consistent with our earlier findings, classification discrepancies remain evident: spaxels assigned to a particular category based on the [N II]-BPT diagram frequently exhibit different classifications when evaluated with the other two diagrams. This persistent inconsistency highlights that classification uncertainties persist regardless of the chosen demarcation approach.

The BPT diagrams in Figure \ref{fig:bpts_law} illustrate the classification of spaxels into distinct categories: SF (blue dots), Sy (orange dots), LI(N)ER (purple dots), and Int (green dots). The classification boundaries shown in magenta are defined by \citet{law2021_refine}. The upper panels (a–c) display all spaxels with a S/N greater than 5. The lower panels (d–f) highlight only those spaxels consistently classified in the same category across all three BPT diagrams. Approximately 11\% of spaxels changed categories across the diagrams and are thus excluded from the lower panels. This exclusion emphasizes the classification uncertainties previously discussed. Table \ref{tab:pure_law} provides a summary of these consistently classified spaxels as illustrated in the lower panels of Figure \ref{fig:bpts_law}.

Based on these analyses, it is clear that previously established demarcation schemes do not ensure consistent classification across all BPT diagrams. For the remainder of our discussion, we will rely on the classification schemes provided by \cite{kewley2001, kewley2006} and \citet{kauffm2003}, as these are the most widely used standards for the classification of optical spectra.

\begin{deluxetable*}{cc|cc|cc}
\tablenum{1}
\tablecaption{Percentage of SF, Comp, AGN, Sy, and LI(N)ERs Classified According to BPT Diagram Boundaries Defined by \citet{kewley2001, kewley2006} and \citet{kauffm2003}} \label{tab:bpts}
\vspace{1cm}%
\tablewidth{0pt}
\tablehead{
\multicolumn{2}{c}{[N II]-BPT} & \multicolumn{2}{c}{[S II]-BPT} & \multicolumn{2}{c}{[O I]-BPT}
}
\startdata
{} & {} & SF & 1 068 283 (97.8\%) & SF & 933 714 (85.4\%)\\
SF & 1 092 832 (84.9\%) & Sy & 9 716 (0.9\%) & Sy & 80 860 (7.4\%) \\
{} & {} & LI(N)ER & 14 833 (1.3\%) & LI(N)ER & 78 258 (7.2\%)\\
\cline{1-6}
{} & {} & SF & 4 454 (6.0\%) & SF & 2 117 (2.8 \%) \\
AGN & 74 589 (5.8\%) & Sy & 25 605 (34.3\%) & Sy & 31 856 (42.7\%)\\
{} & {} & LI(N)ER & 44 530 (59.7\%) & LI(N)ER & 40 616 (54.5\%) \\
\cline{1-6}
{} & {} & SF & 92 837 (77.9\%) & SF & 61 261 (51.4\%) \\
Comp & 119 222 (9.3\%) & Sy & 2 386 (2.0\%) & Sy & 7 133 (6.0\%) \\
{} & {} & LI(N)ER & 23 999 (20.1\%) & LI(N)ER & 50 828 (42.6\%) \\
\enddata
\tablecomments{The percentages in the [S II]- and [O I]-BPT diagrams are calculated relative to the number of spaxels classified within each corresponding category of the [N II]-BPT diagram (first column), rather than the total number of spaxels. Only percentages in the [N II]-BPT diagram reflect the total spaxel count used in the analysis. Classification boundaries are derived from \citet{kewley2001, kewley2006} and \citet{kauffm2003}.}
\end{deluxetable*}

\begin{deluxetable*}{cc|cc|cc}
\tablenum{2}
\tablecaption{Percentage of SF, Int, AGN, Sy, and LI(N)ERs Classified According to BPT Diagram Boundaries Defined by \citet{law2021_refine}} \label{tab:bpts_law}
\vspace{1cm}%
\tablewidth{0pt}
\tablehead{
\multicolumn{2}{c}{[N II]-BPT} & \multicolumn{2}{c}{[S II]-BPT} & \multicolumn{2}{c}{[O I]-BPT}
}
\startdata
{} & {} & SF & 1 108 183 (99.6\%) & SF & 1 094 585 (98.4\%)\\
SF & 1 112 655 (86.5\%) & Sy & 126 (0.01\%) & Sy & 1 622 (0.15\%) \\
{} & {} & LI(N)ER & 690 (0.06\%) & LI(N)ER & 3 841 (0.35\%)\\
{} & {} & Int & 3 656 (0.33\%) & Int & 12 607 (1.13\%)\\
\cline{1-6}
{} & {} & SF & 1 000 (4.52\%) & SF & 1 088 (4.91 \%) \\
Sy & 22 146 (1.72\%) & Sy & 18 336 (82.8\%) & Sy & 19 728 (89.1\%)\\
{} & {} & LI(N)ER & 1 603 (7.24\%) & LI(N)ER & 1 002 (4.52\%) \\
{} & {} & Int & 1 207 (5.45\%) & Int & 328 (1.48\%) \\
\cline{1-6}
{} & {} & SF & 15 651 (24.5\%) & SF & 15 891 (24.9\%) \\
LI(N)ER & 63 799 (4.96\%) & Sy & 2 334 (3.66\%) & Sy & 5 909 (9.26\%) \\
{} & {} & LI(N)ER & 34 765 (54.5\%) & LI(N)ER & 29 901 (46.9\%) \\
{} & {} & Int & 11 049 (17.3\%) & Int & 12 098 (19.0\%) \\
\cline{1-6}
{} & {} & SF & 77 311 (87.8\%) & SF & 72 686 (82.5\%) \\
Int & 88 055 (6.84\%) & Sy & 329 (0.37\%) & Sy & 718 (0.82\%) \\
{} & {} & LI(N)ER & 4 103 (4.66\%) & LI(N)ER & 5 184 (5.89\%) \\
{} & {} & Int & 6 312 (7.17\%) & Int & 9 467 (10.8\%) \\
\enddata
\tablecomments{Similar to Table \ref{tab:bpts}, but percentages are calculated using the BPT diagram classification boundaries defined by \citet{law2021_refine}.}
\end{deluxetable*}

\begin{deluxetable}{l|c}
\tablenum{3}
\tablecaption{Number of pure samples and their percentages defined by \citet{kewley2001, kewley2006} and \citet{kauffm2003} \label{tab:pure}}
\vspace{1cm}%
\tablewidth{0pt}
\tablehead{
{Name} & {Number}}
\startdata
Classified as SF in all BPT diagrams & 931 887 (72.4\%) \\
AGN in [N II]-BPT \& Sy in the other two & 24 320 (1.89\%) \\
AGN in [N II]-BPT \& LI(N)ERs in the other two & 38 097 (2.96\%) \\
Comp spaxels in [N II]-BPT & 173 129 (13.5 \%)\\
Ambiguous spaxels & 119 222 (9.27\%) \\
\hline
Total number of spaxels after S/N cuts & 1 286 655 (100\%) \\
\enddata
\end{deluxetable}

\begin{deluxetable}{l|c}
\tablenum{4}
\tablecaption{Number of pure samples and their percentages defined by \citet{law2021_refine}} \label{tab:pure_law}
\vspace{1cm}%
\tablewidth{0pt}
\tablehead{
{Name} & {Number}}
\startdata
Classified as SF in all BPT diagrams & 1 093 599 (85.0\%) \\
Classified as Sy in all BPT diagrams & 17 948 (1.39\%) \\
Classified as LI(N)ER in all BPT diagrams & 25 256 (1.96\%) \\
Classified as Int in all BPT diagrams & 3 062 (0.24 \%)\\
Ambiguous spaxels & 146 790 (11.4\%) \\
\hline
Total number of spaxels after S/N cuts & 1 286 655 (100\%) \\
\enddata
\end{deluxetable}

\subsubsection{Testing set (Ambiguous data)}
\label{subsec:amb}

For testing the trained model, we incorporate the ambiguous spaxels, which are not consistently classified and thus difficult to assign definitively to any specific class. Our analysis identified a total of 119 222 such spaxels. The goal is to evaluate the model’s ability to distinguish and classify ambiguous data that did not clearly belong to a single category, providing a rigorous test of its robustness and accuracy. By analyzing how the UMAP-trained model classifies these ambiguous data, we aim to gain insights into their behavior and underlying properties. Examining their distribution and clustering within the model's classification scheme helps us better understand their characteristics and can guide future improvements.%

\section{Dimensionality Reduction and Clustering}
\label{sec:model}

\subsection{Dimensionality Reduction}

We employ the UMAP algorithm for its flexibility, computational efficiency, and ability to capture global structures. UMAP constructs a high-dimensional graph that encodes relationships between data points and then optimizes a low-dimensional representation that preserves the original structure. Compared to traditional methods like principal component analysis (PCA), UMAP excels in handling non-linear relationships and complex manifold structures, making it highly effective for applications such as clustering, data visualization, and noise reduction.

The BPT diagrams are based on four emission-line ratios---[O III]$\lambda$5008/H$\beta$, [N II]$\lambda$6585/H$\alpha$, [S II]$\lambda\lambda$6718,32/H$\alpha$, and [O I]$\lambda$6302/H$\alpha$---creating a four-dimensional space; however, we typically rely on three separate diagrams, each focusing on a pair of emission line ratios. This approach provides only a limited view of the full parameter space, which may obscure the underlying complexities within the data. To address this, we applied UMAP to project these four dimensions (4D) into a 2D space, enhancing classification clarity and revealing hidden substructures. By combining all four parameters into a 2D representation, UMAP visually separates distinct classes and reveals patterns that might remain concealed in higher-dimensional spaces.

In UMAP, three primary hyperparameters significantly influence the embedding process: \texttt{n\_neighbors}, \texttt{min\_dist}, and \texttt{n\_components}. The \texttt{n\_neighbors} parameter determines the size of the local neighborhood used in manifold approximation. The \texttt{n\_neighbors} parameter defines the number of nearest neighbors considered when approximating the manifold structure. Smaller values emphasize local relationships, whereas larger values help capture global data structure. For this study, we selected \texttt{n\_neighbors} = 1 000 to effectively identify global structural features. The \texttt{min\_dist} parameter sets the minimum allowed distance between embedded points, influencing the compactness and clustering in the resulting embedding. Lower \texttt{min\_dist} values produce denser, more detailed local clusters, while higher values result in embeddings with broader, less dense structures. In this analysis, we employed \texttt{min\_dist} = 0.0 to maximize clustering clarity. Lastly, the \texttt{n\_components} parameter determines the dimensionality of the embedding space; we set \texttt{n\_components} = 2 for clear visualization and straightforward interpretation.

To train the UMAP model, we used the four traditional BPT emission-line ratios: $\log([\text{O III}]\lambda5008/\text{H}\beta)$, $\log([\text{N II}]\lambda6585/\text{H}\alpha)$, $\log([\text{S II}]\lambda\lambda6718,32/\text{H}\alpha)$, and $\log([\text{O I}]\lambda6302/\text{H}\alpha)$. These ratios are derived from emission lines close in wavelength, making them largely independent of reddening effects. We selected these particular emission-line ratios not only due to their established usage in the literature but also because they effectively distinguish between ionization sources, providing clear and physically meaningful separation. After testing various combinations of emission lines, line ratios, and hyperparameters, the chosen set of four ratios consistently yielded the most distinct classification results. The UMAP algorithm was implemented using the \texttt{Python} package \texttt{umap-learn} \citep{umap}. Training and testing were conducted using the clean and ambiguous datasets as described in Section \ref{sec:sample}.

\subsection{Clustering}

Upon completing the training phase, we applied the HDBSCAN (Hierarchical Density-Based Spatial Clustering of Applications with Noise) algorithm \citep{hdbscan} to cluster the dataset, which had been projected into a lower-dimensional space using the UMAP model. HDBSCAN, a density-based clustering technique, is advantageous as it does not require the number of clusters to be predefined. For our analysis, we set \texttt{min\_samples} to 10, defining the minimum number of points required to form a dense region, and \texttt{min\_cluster\_size} to 500, ensuring that clusters smaller than this threshold are classified as noise. This approach allows for the identification of clusters with varying densities and shapes while effectively filtering out noise, providing enhanced flexibility and precision in the analysis of complex data structures. The HDBSCAN algorithm used in this analysis is implemented via the \texttt{Python} package \texttt{hdbscan} \citep{hdbscan}.

\section{Results}
\label{sec:results}

\subsection{The UMAP training and testing results}

\begin{figure}
    \centering
    \includegraphics[width=0.45\textwidth]{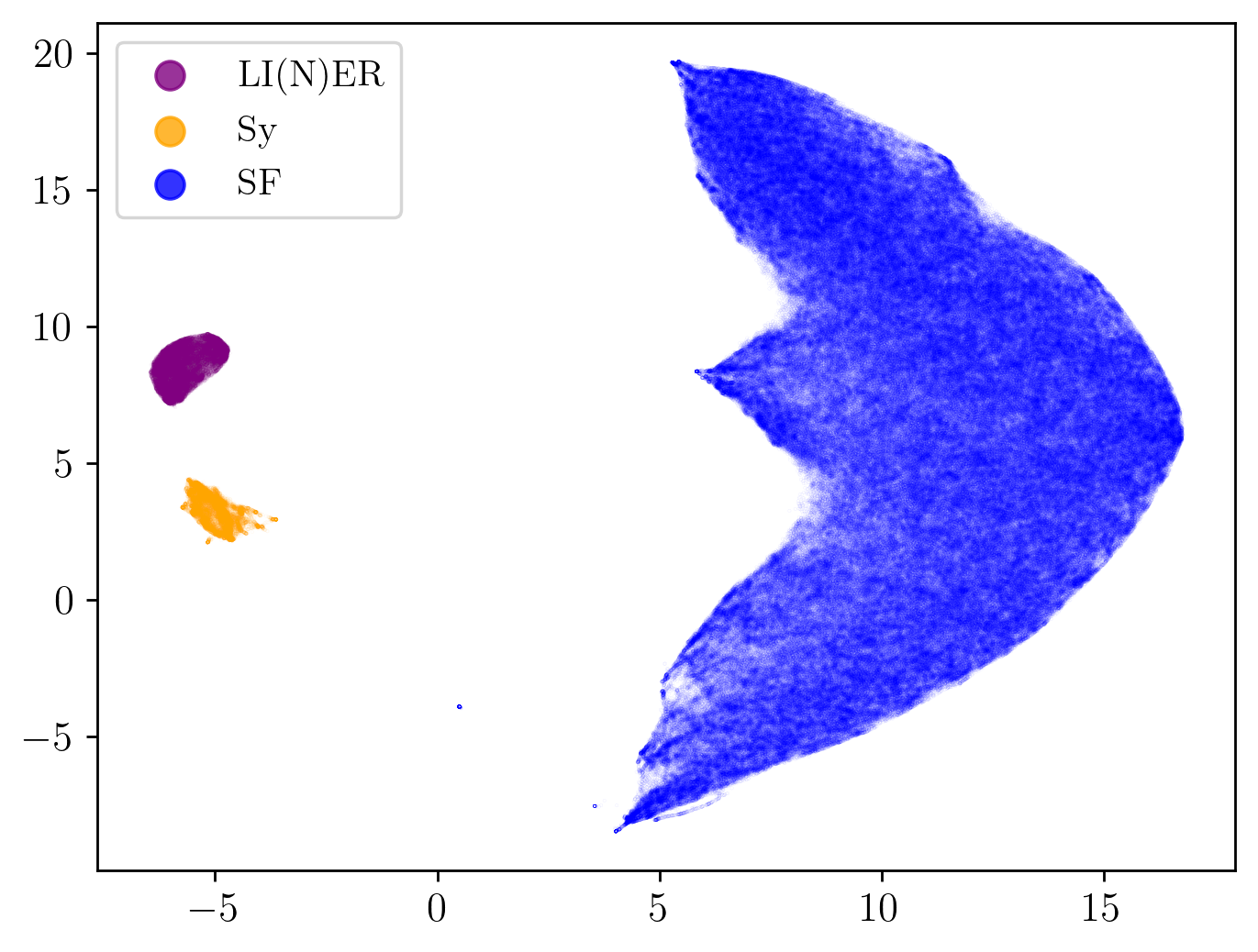}
    \includegraphics[width=0.45\textwidth]{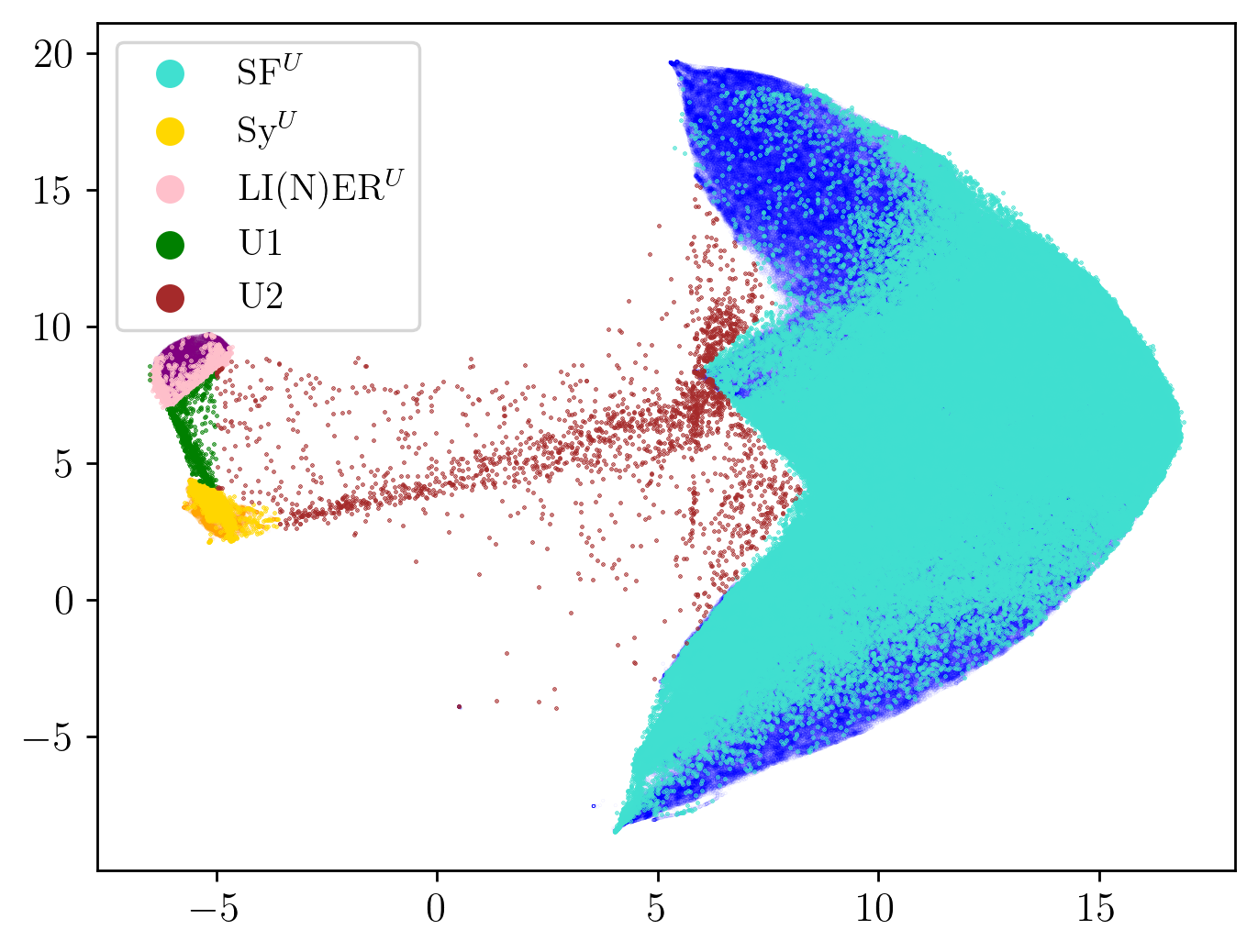}
    \caption{UMAP Embeddings for consistently classified spaxels (Top), ambiguous spaxels (Bottom)}
    \label{fig:umap_combined}
\end{figure}

\begin{deluxetable}{lc}
\tablenum{5}
\tablecaption{Number of ambiguous samples classified by the model during analysis
\label{tab:ambi}}
\tablewidth{0pt}
\tablehead{
{Name} & {Number}}
\startdata
SF$^U$  & 101 527 (85.16\%) \\
Sy$^U$ & 4 989 (4.18\%) \\
LI(N)ER$^U$ & 3 989 (3.35\%) \\
Unclassified 1 (U1) & 703 (0.59\%) \\
Unclassified 2 (U2) & 8 014 (6.72\%) \\
\hline
Total & 119 222 (100\%) \\
\enddata
\end{deluxetable}

We developed the unsupervised machine learning model using the training and testing sets outlined in Section \ref{sec:sample}. For dimensionality reduction, classification, and clustering, we applied the UMAP algorithm alongside the HDBSCAN method, as detailed in Section \ref{sec:model}. Figure \ref{fig:umap_combined}, shown at the top of the figure, presents the 2D embeddings generated by the UMAP model trained on the clean dataset from Section \ref{subsec:clean}. In this figure, the blue, orange, and purple dots represent SF, Sy, and LI(N)ER spaxels, respectively, as consistently classified by the three BPT diagrams. The distinct separation between these classes, especially visible on the upper side of Figure \ref{fig:umap_combined}, highlights the model's ability to accurately distinguish between them.

In addition, we tested the model using the ambiguous spaxels defined in Section \ref{subsec:amb} to examine their distribution within the 2D embedding space produced by the UMAP model. The bottom panel of Figure \ref{fig:umap_combined} displays the results of the testing set (ambiguous spaxels) overlaid on the training set, which consists of the clean classification spaxels. For further analysis, we classified these points based on their positions in the UMAP space. Specifically, SF$^U$ (turquoise dots), Sy$^U$ (gold dots), and LI(N)ER$^U$ (pink dots) denote the ambiguous data points falling within the previously defined SF, Sy, and LI(N)ERs, respectively, indicating they are newly classified as SF, Sy, or LI(N)ER---data that the traditional BPT diagrams were unable to differentiate.

Some spaxels, however, do not fall within any of the predefined UMAP clusters. To further investigate their properties, we categorized this unclassified group into two subgroups: U1 (dark green dots) and U2 (brown dots). The U1 points are positioned closer to the AGN region, while U2 points lie between the AGN and SF regions. Table \ref{tab:ambi} summarizes the number of ambiguous spaxels newly classified using the updated UMAP classification method. These results help quantify how many previously ambiguous spaxels can now be systematically categorized.

\subsection{The UMAP-classified data on the BPT diagrams}

\begin{figure*}
\centering
\includegraphics[width=0.75\textwidth]{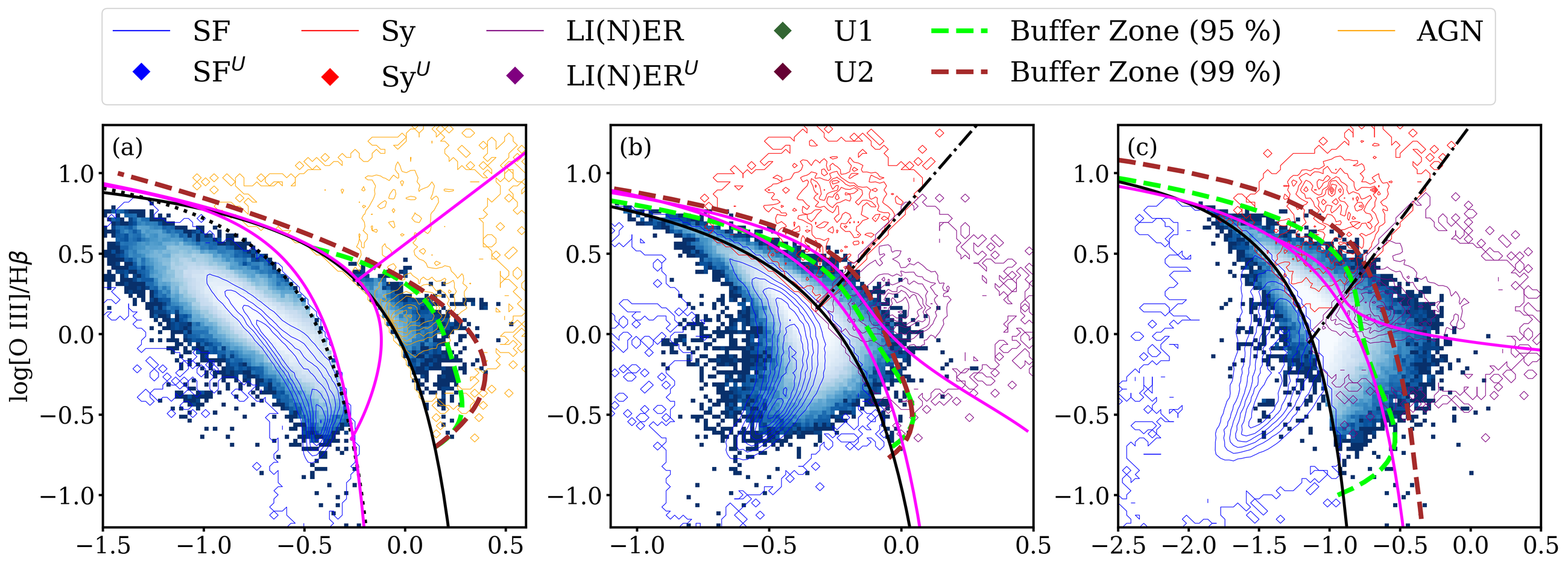}
\includegraphics[width=0.75\textwidth]{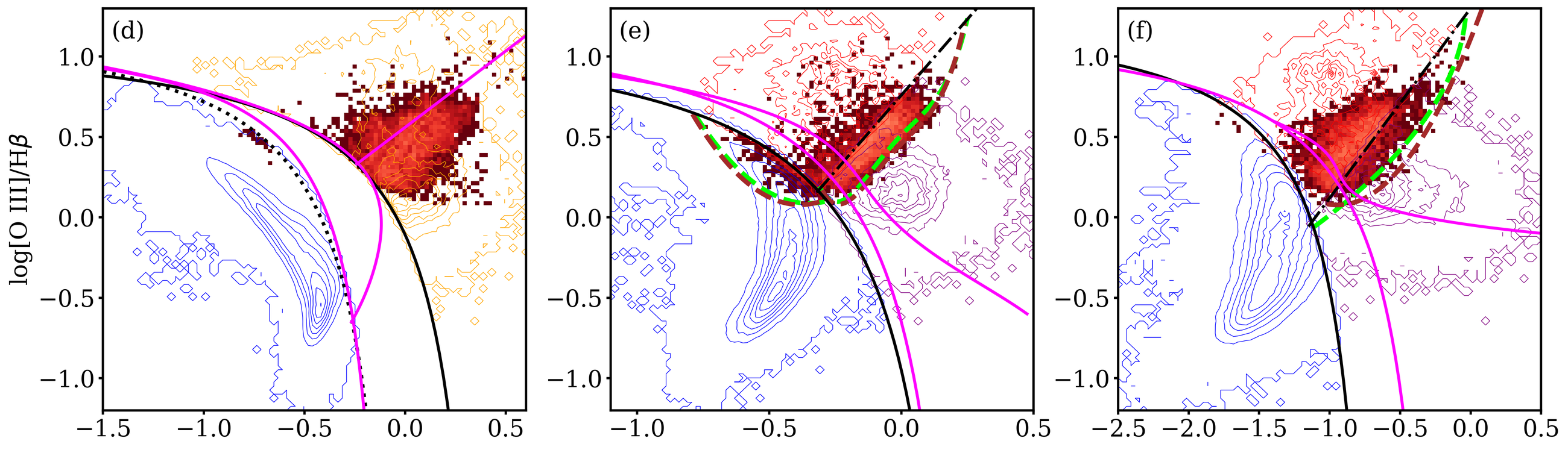}
\includegraphics[width=0.75\textwidth]{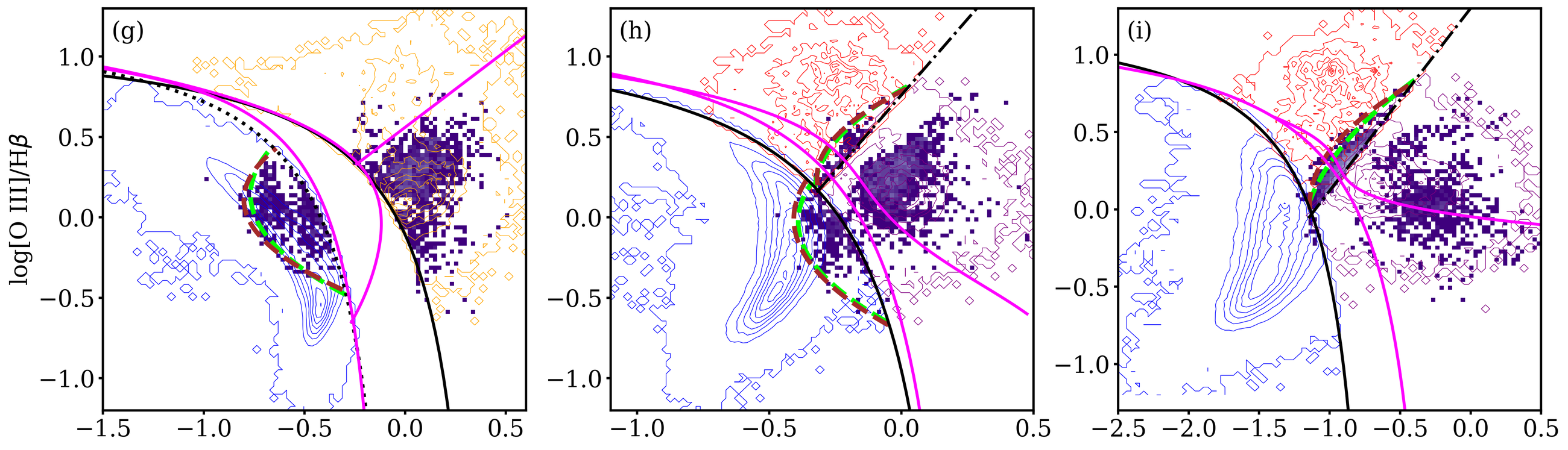}
\includegraphics[width=0.75\textwidth]{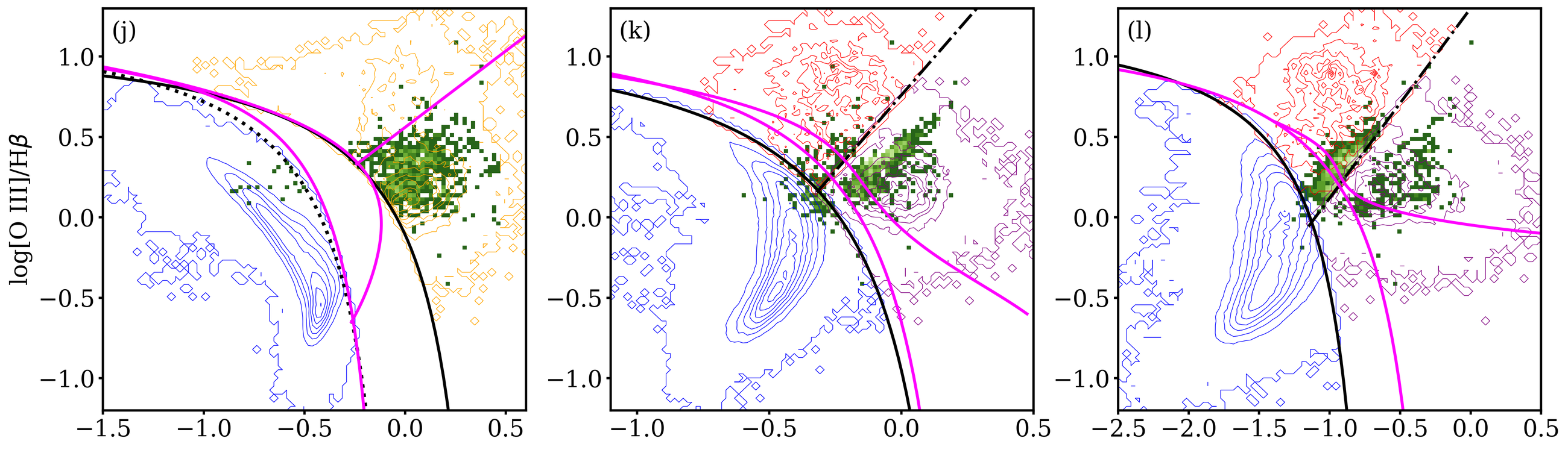}
\includegraphics[width=0.75\textwidth]{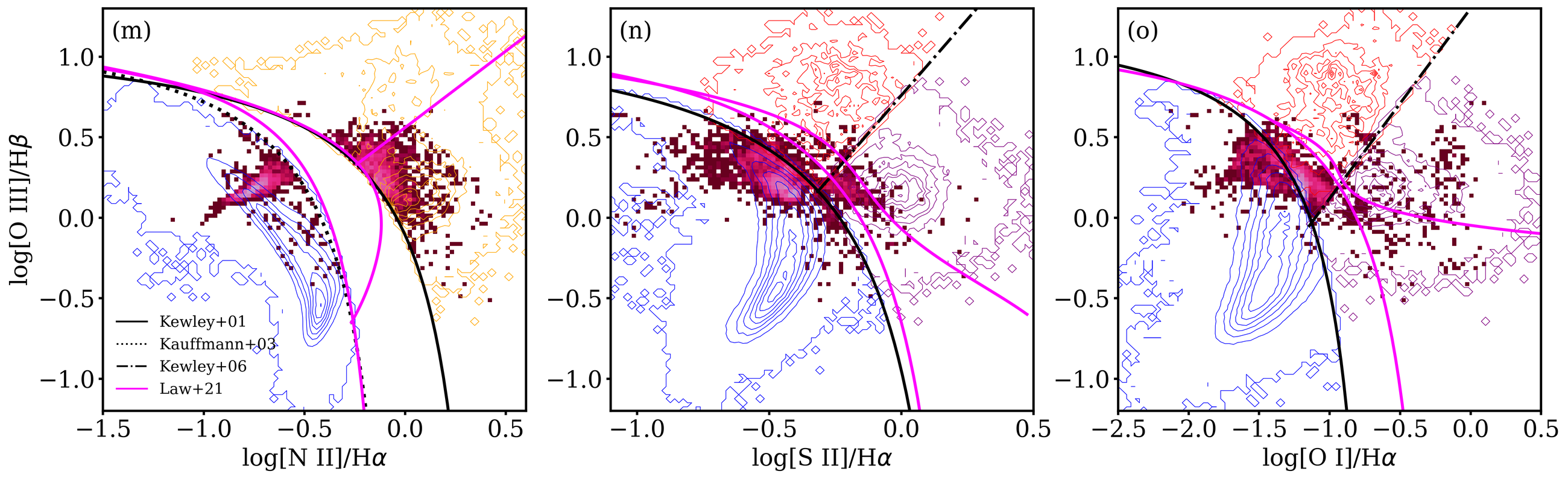}
\caption{The three BPT diagrams illustrating the distribution of ambiguous spaxels classified by UMAP into SF$^U$ (first row), Sy$^U$ (second row), LI(N)ER$^U$ (third row), as well as ambiguous spaxels that remain unclassified, labeled as U1 (fourth row) and U2 (fifth row). The dashed curves indicate the buffer zones newly defined by the UMAP classification analysis. The lime dashed curves enclose 95\% of the spaxels, while the brown dashed curves enclose 99\%.}
\label{fig:umap_amb_bpt}
\end{figure*}

To investigate the behavior of spaxels newly classified by the UMAP method from previously ambiguous data, we plotted these points on the BPT diagrams to examine their distribution relative to the original clean dataset. Figure \ref{fig:umap_amb_bpt} display the UMAP-classified categories SF$^U$, Sy$^U$, and LI(N)ER$^U$, along with two unclassified subclasses, U1 and U2. This comparison provides insight into the distribution differences and highlights the distinctions introduced by the updated UMAP classifications.

In Figure \ref{fig:umap_amb_bpt}, we observe the distributions of UMAP-classified spaxels: SF$^U$ (blue), Sy$^U$ (red), and LI(N)ER$^U$ (purple). To compare with the clean classification data derived from the traditional BPT diagrams, we overlay these distributions with contours on each BPT diagram. Note that the color scheme of the 2D histogram is inverted, meaning that the white regions within the distributions indicate areas with the highest concentration of data points. This visualization helps to highlight the overlap and distinctions between the UMAP-classified data and the traditional BPT classification.

It is evident that the UMAP-classified categories SF$^U$, Sy$^U$, and LI(N)ER$^U$ are not strictly confined within the boundaries established by the \citet{kewley2006} and \citet{kauffm2003} curves. Across all BPT diagrams, we observe that the previous demarcations are crossed, with some spaxels appearing in regions designated for other classes. This discrepancy underscores that current demarcation curves do not consistently classify all observed spectra. By using a UMAP classification scheme trained on "clean" spaxels to classify spectra with ambiguous classification, we get a complementary perspective regarding their nature. This suggests that machine learning approaches such as UMAP can serve as a useful tool to improve current classification techniques. For spaxels located outside their previously defined boundaries, we introduce lime and brown dashed curves to denote the newly defined buffer regions that enclose 95\% and 99\% of the spaxels, respectively, measured outward from the predetermined demarcation curves. These curves help visualize the distribution and concentration of spaxels relative to the original boundaries. A detailed description of the boundary equations and their implementation is provided in Appendix \ref{app:boundary}.

The resulting boundaries are derived from the multidimensional UMAP classification space rather than from the 2D BPT projections alone, and therefore reflect the statistical spread of spectra with similar characteristics rather than a simple horizontal or vertical shift of the traditional demarcation lines. Notably, the buffer regions for SF$^U$ spaxels on the [S II]- and [O I]-BPT diagrams show partial agreement with the \citet{law2021_refine} curves, indicating that both approaches capture a similar trend at the SF-AGN transition. For the Sy$^U$ and LI(N)ER$^U$ classes, the buffer regions reveal a smoother transition rather than a sharp division, consistent with a physical blending of ionization conditions instead of a fixed offset. This underscores the complementary role of UMAP-based classification, offering a quantitative measure of uncertainty and overlap in spectral parameter space.

Figure \ref{fig:umap_amb_bpt} also illustrates the distribution of data unclassified by the UMAP model across the BPT diagrams. The fourth row of panels represent the U1 class, which, although unclassified, is located near the AGN clusters on the bottom of the Figure \ref{fig:umap_combined}. These spaxels exhibit both Sy and LI(N)ER characteristics. Most of the U1 points fall within the AGN region of the [N II]-BPT diagram, but on the [S II]- and [O I]-BPT diagrams, they predominantly occupy the Sy and LI(N)ER regions, respectively. The bottom panels display the U2 class, also unclassified, but situated between the AGN and SF regions. U2 data points are distributed more broadly across the BPT diagrams; while the majority are located in the AGN region on the [N II]-BPT diagram, most points fall within the SF region on the [S II]-BPT diagram. The distribution on the [O I]-BPT diagram is more challenging to interpret. This highlights the complexity of classifying such data using the traditional demarcation curves, underscoring the limitations of the previous classification schemes.

\subsection{Enhancing classification and gaining deeper insights through additional properties}

As shown in Table \ref{tab:bpts}, our statistical analysis also highlighted inconsistencies in traditional BPT diagnostics across the three diagrams, sometimes leading to misclassifications. We show that UMAP can address these limitations and serve as a valuable supplement to BPT diagram classification. Incorporating additional physical properties enables a deeper understanding of spaxels, particularly those with ambiguous classifications. Previous studies have taken similar approaches by including emission line strengths \citep[e.g., H$\alpha$ flux;][]{dagostino2019} and kinematic information \citep[e.g., H$\alpha$ velocity dispersion;][]{law2021_refine, RN170} to refine and distinguish between different ionization sources. Following this line of work, we aim to apply comparable diagnostics to our newly classified ambiguous spaxels in order to investigate their physical nature and uncover patterns in their spatial and spectral behavior.

Even with just the four BPT emission line ratios, we gained valuable insights by applying the UMAP algorithm, which helped uncover meaningful patterns and classifications within the spaxels. By incorporating additional parameters---such as spatial, kinetic information like velocity dispersion of emission lines, and intrinsic luminosities \citep{RN170}---we can further enhance this analysis. These additional properties provide a more comprehensive and detailed understanding of the spaxels. To visualize these improvements, we incorporated new parameters into our UMAP-classified spaxels and generated color-coded plots. To ensure the robustness of the results, we excluded the outermost 5$\%$ of the data from both the lower and upper extremes to minimize the influence of outliers.

\subsubsection{Normalized elliptical radius}

\begin{figure*}
\plotone{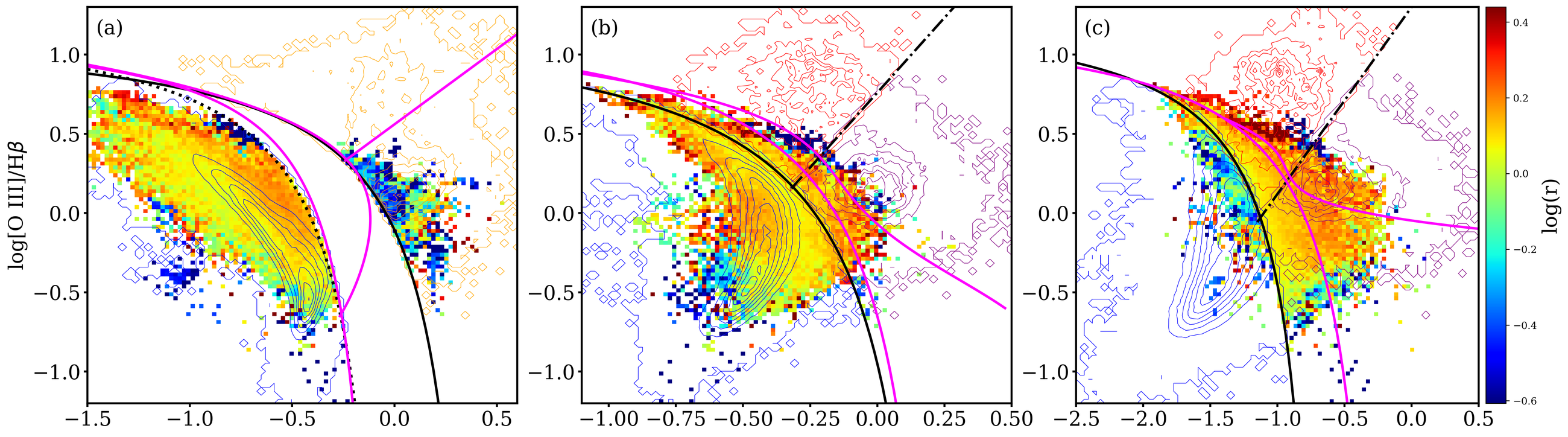}
\plotone{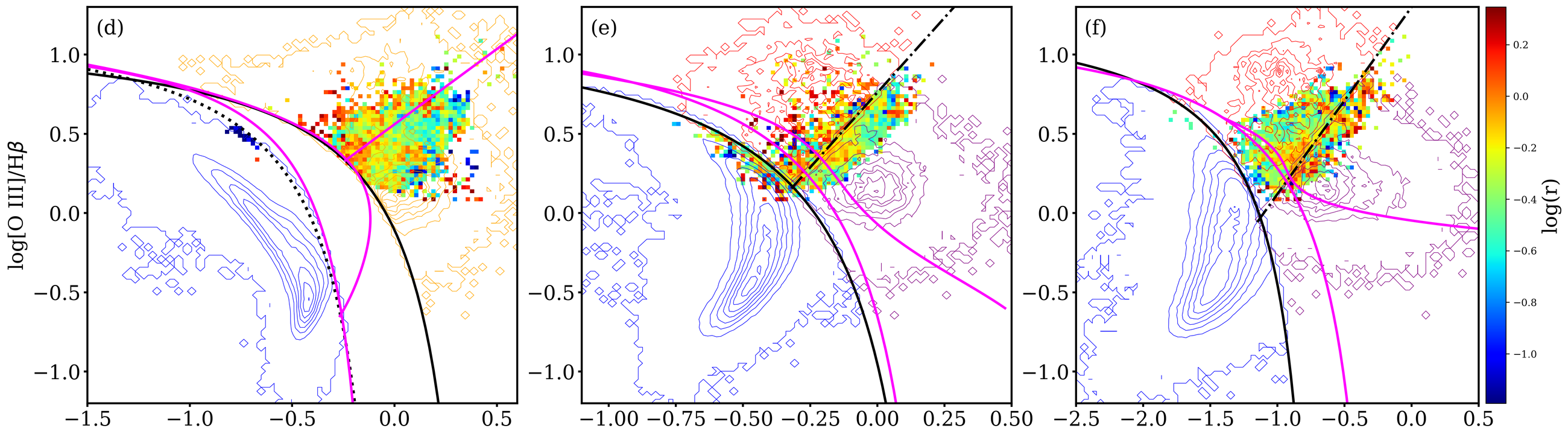}
\plotone{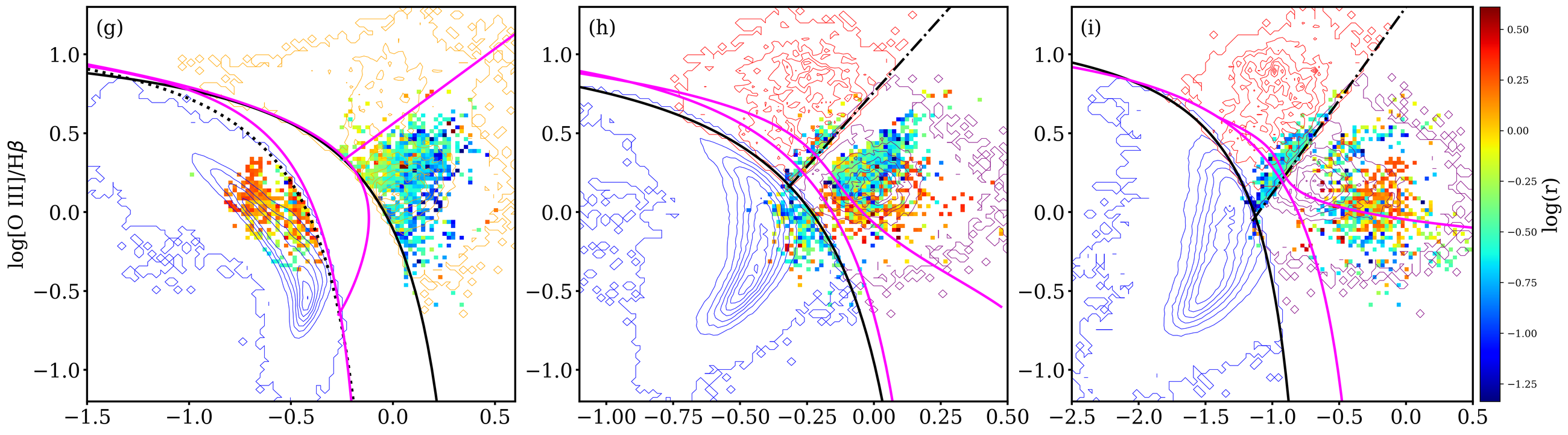}
\plotone{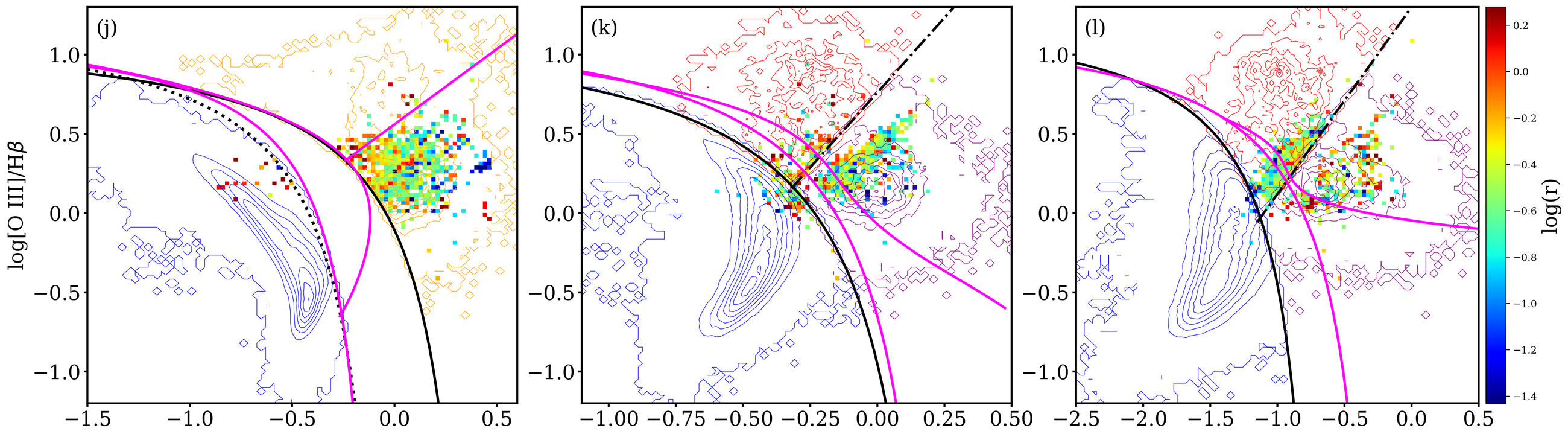}
\plotone{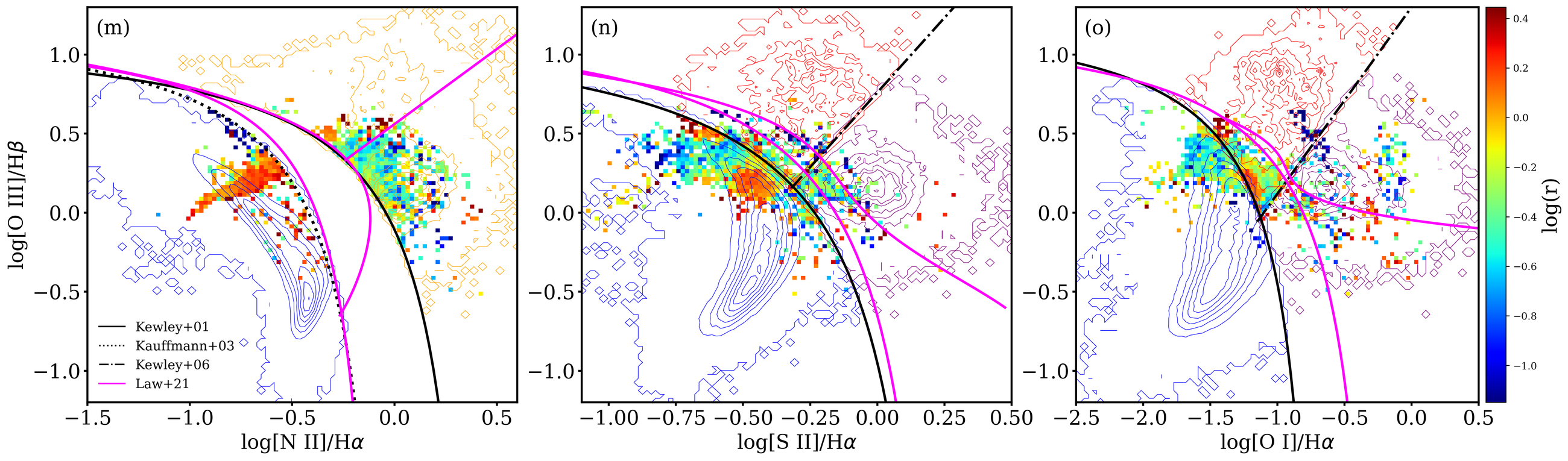}
\caption{Color-coded plots showing the normalized elliptical radius (\texttt{spx\_ellcoo\_Reff}) for the five UMAP-classified categories: SF$^U$ (top row), Sy$^U$ (second row), LI(N)ER$^U$ (third row), U1 (fourth row), and U2 (bottom row).} \label{fig:umap_amb_bpt_r}
\end{figure*}

We characterize the radial behavior of spaxels using the normalized elliptical radius (\texttt{spx\_ellcoo\_Reff} parameter provided by the MaNGA), which expresses the galactocentric distance in units of the effective radius. Figure \ref{fig:umap_amb_bpt_r} presents color-coded plots based on the normalized elliptical radius for the SF$^U$ and LI(N)ER$^U$ clusters on the BPT diagrams. To improve clarity and reduce the impact of outliers, we first excluded the outer 5\% of spaxels based on their property distributions. We then applied 2D spatial binning and assigned each bin a color corresponding to the median normalized elliptical radius of the spaxels it contains.

For the SF$^U$ cluster (top row), a clear radial gradient is visible on the [N II]-BPT diagram: spaxels on the AGN-side are located closer to the galactic center (bluer), while those in the SF region are more extended. In contrast, the trend reverses in the [O I]-BPT diagram, where SF-side spaxels are more centrally located. No distinct trend appears on the [S II]-BPT diagram.

Figure \ref{fig:umap_amb_bpt_r} (third row) reveals a notable dichotomy within the LI(N)ER$^U$ population on the [N II]-BPT diagram: some spaxels fall on the SF-side of the demarcation line (likely LIERs), while others lie within the AGN region (likely LINERs). This distinction is supported by narrower H$\alpha$ line widths (Figure \ref{fig:umap_amb_bpt_ha}) and lower [O III] luminosities (Figure \ref{fig:umap_amb_bpt_o3}) for the former group. While LIERs and LINERs largely overlap on the [S II]-BPT, the [O I]-BPT shows tentative separation, with LIERs occupying regions of higher [O I]/H$\alpha$ ratios. These trends suggest that the [N II]-BPT diagram, especially when combined with kinematic and spatial information, may offer a useful way to distinguish between LIERs and LINERs.

Sy$^U$ and U1 clusters (second and fourth rows) show no clear radial trends across any of the BPT diagrams, suggesting these groups may be more mixed in physical origin or not strongly structured in radius.

For the U2 cluster (bottom row), a radial gradient appears on the [N II]-BPT diagram, with spaxels in the AGN region being slightly more central than those on the SF-side. In particular, the [O I]-BPT diagram (panel (o)) shows a deviation that likely stems from the use of the earlier [O I] demarcation line from \citet{kewley2001}. Applying the revised boundary from \citet{law2021_refine} would classify the majority of these spaxels as SF, bringing the results into clear agreement with the [N II]- and [S II]-BPT diagrams.

\subsubsection{H$\alpha$ velocity dispersion}
\begin{figure*}
\plotone{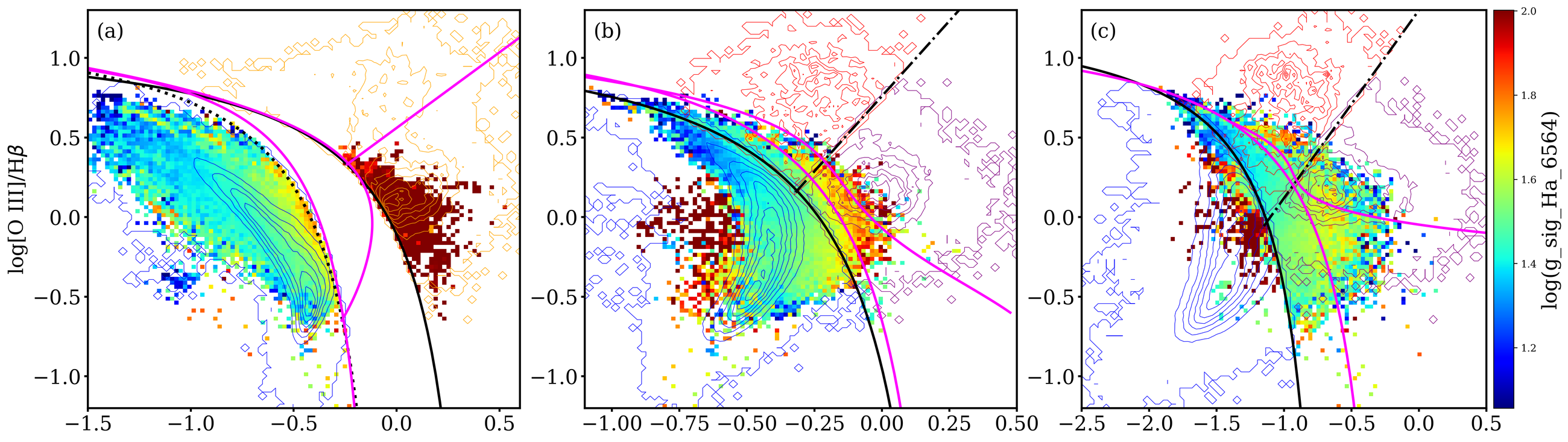}
\plotone{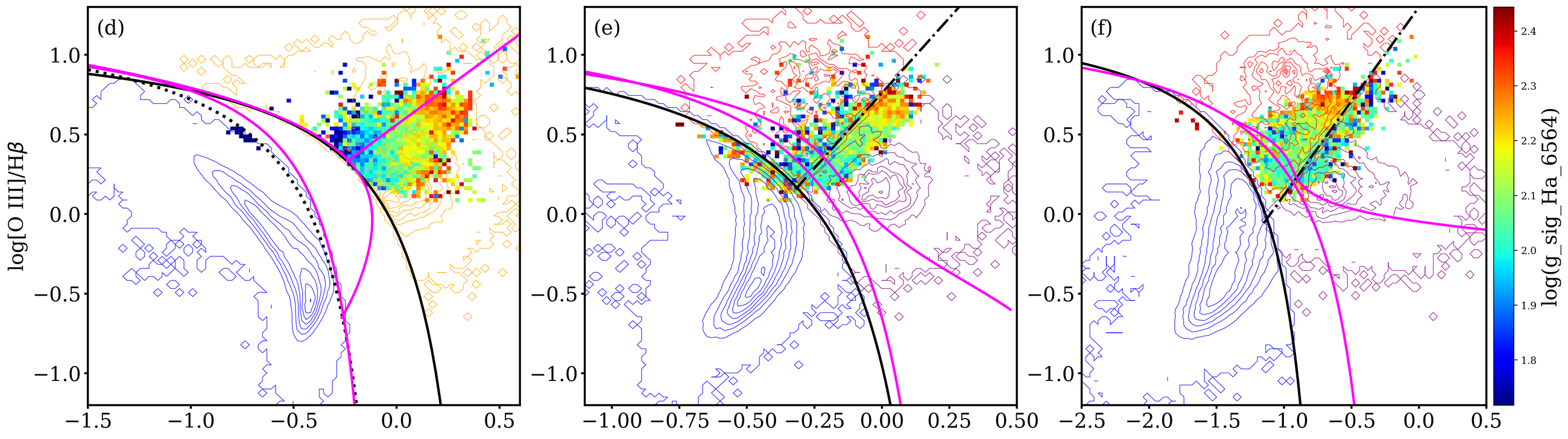}
\plotone{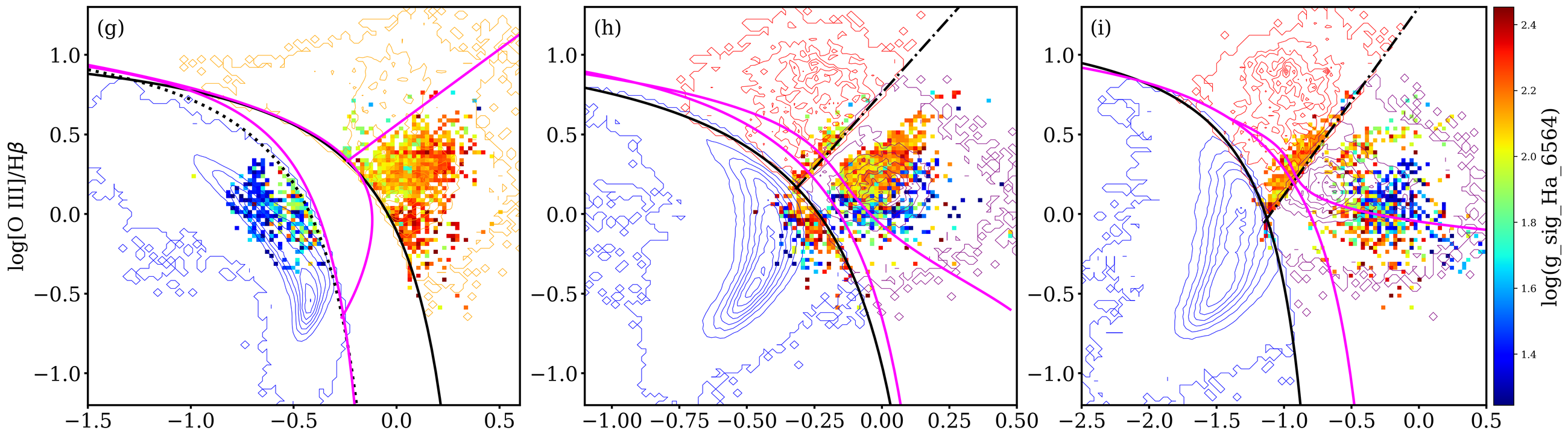}
\plotone{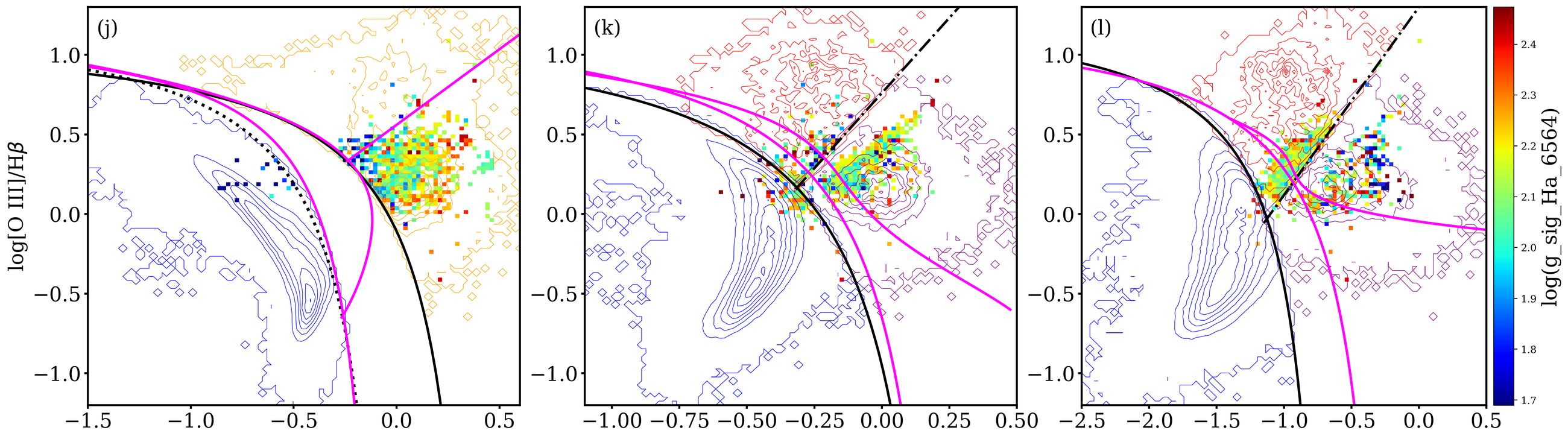}
\plotone{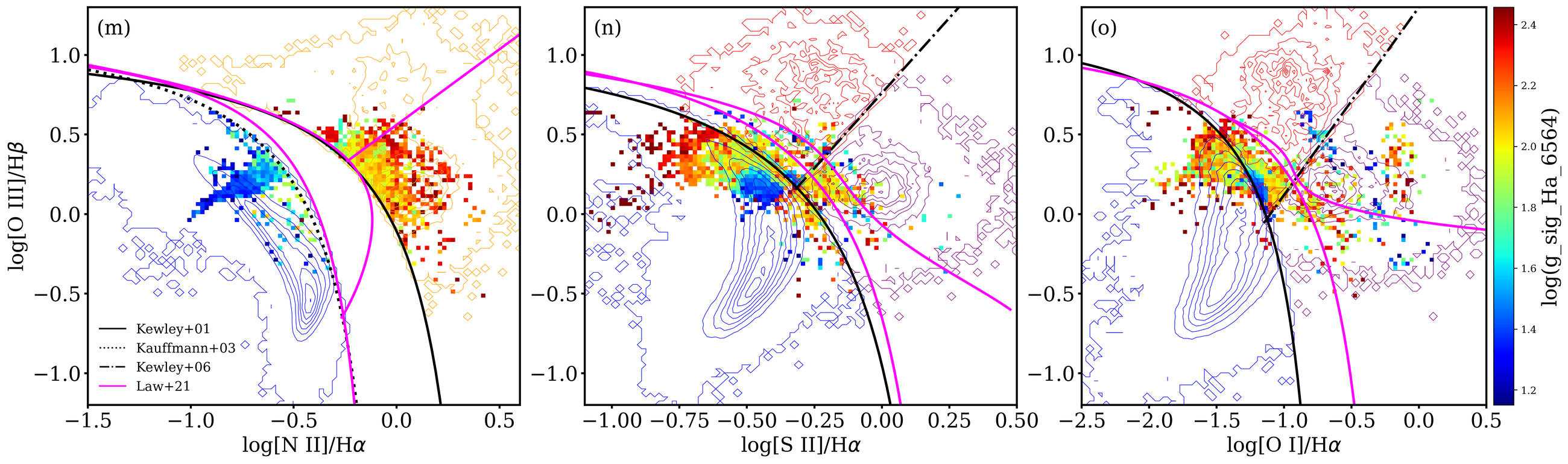}
\caption{Color-coded plots showing the velocity dispersion of the H$\alpha$ emission line for the five UMAP-classified categories: SF$^U$ (top row), Sy$^U$ (second row), LI(N)ER$^U$ (third row), U1 (fourth row), and U2 (bottom row).}
\label{fig:umap_amb_bpt_ha}
\end{figure*}

Figure \ref{fig:umap_amb_bpt_ha} presents color-coded BPT diagrams for the five UMAP-classified clusters---SF$^U$, Sy$^U$, LI(N)ER$^U$, U1, and U2---where each 2D bin is shaded by the median velocity dispersion of the H$\alpha$ emission line in log scale. These maps offer insight into the kinematic structure of the ionized gas, which reflects underlying energetic processes such as AGN-driven outflows, supernova feedback, galactic winds, and shocks. While AGN and starburst activity can both influence velocity dispersions, shocks are especially relevant in the context of LI(N)ER emission and are known to broaden line profiles significantly. Therefore, velocity dispersion can serve as a useful, if indirect, proxy for shock-dominated ionization \citep[e.g.,][]{rich2011, Belfiore2016}.

For the SF$^U$ cluster (top row), the spaxels generally exhibit relatively lower velocity dispersions across all three BPT diagrams, consistent with expectations for SF regions. Notably, velocity dispersion increases slightly toward the AGN boundary, especially on the [N II]-diagram, while some of the higher $\sigma_{H\alpha}$ spaxels located in the SF-side on the [S II]- and [O I]-BPT diagram, hinting at a possible contribution from non-thermal processes near those borders.

The LI(N)ER$^U$ cluster (third row) displays a internal dichotomy, most clearly in the [N II]-BPT diagram: some spaxels lie on the SF-side with relatively lower dispersions ($\log(\sigma_{H\alpha}) \lesssim$ 2.0), while others lie well within the AGN region and exhibit significantly broader H$\alpha$ lines ($\log(\sigma_{H\alpha}) \gtrsim$ 2.0). This division is consistent with previous work showing that increased velocity dispersion often signals shock-dominated ionization or weak AGN activity \citep[e.g.,][]{rich2011, Rich2014}. Accordingly, the lower-dispersion branch is more likely associated with extended LIER emission powered by evolved stellar populations, whereas the higher-dispersion branch is more in line with nuclear LINER-like excitation. While this dichotomy is somewhat blurred on the [S II]-BPT diagram, the [O I]-BPT diagram continues to reveal a dispersion gradient within LI(N)ER$^U$, reinforcing the view that shocks or low-level AGN contribute to the observed emission.

For the Sy$^U$ and U1 clusters (second and fourth rows), no strong or coherent patterns in H$\alpha$ velocity dispersion are evident across most of the BPT diagrams, suggesting comparatively heterogeneous kinematic structures. However, for the Sy$^U$ cluster on the [N II]-BPT diagram, a gradient is visible, with velocity dispersion increasing along the [N II]/H$\alpha$ ratio.

In contrast, the U2 cluster (bottom row) displays a noticeable velocity dispersion gradient, particularly on the [N II]-BPT diagram: spaxels located on the SF-side exhibit lower H$\alpha$ dispersions, while those extending into the AGN region show significantly broader profiles. This gradient becomes less pronounced on the [S II]- and [O I]-BPT diagrams; instead, spaxels with narrow H$\alpha$ velocity dispersions appear to cluster near the central region of the U2 distribution. The spatial and kinematic behavior of U2 may indicate a transitional population influenced by both star-formation and AGN-related processes, warranting further investigation into its ionization mechanisms.

\subsubsection{[O III] luminosity}

\begin{figure*}
\plotone{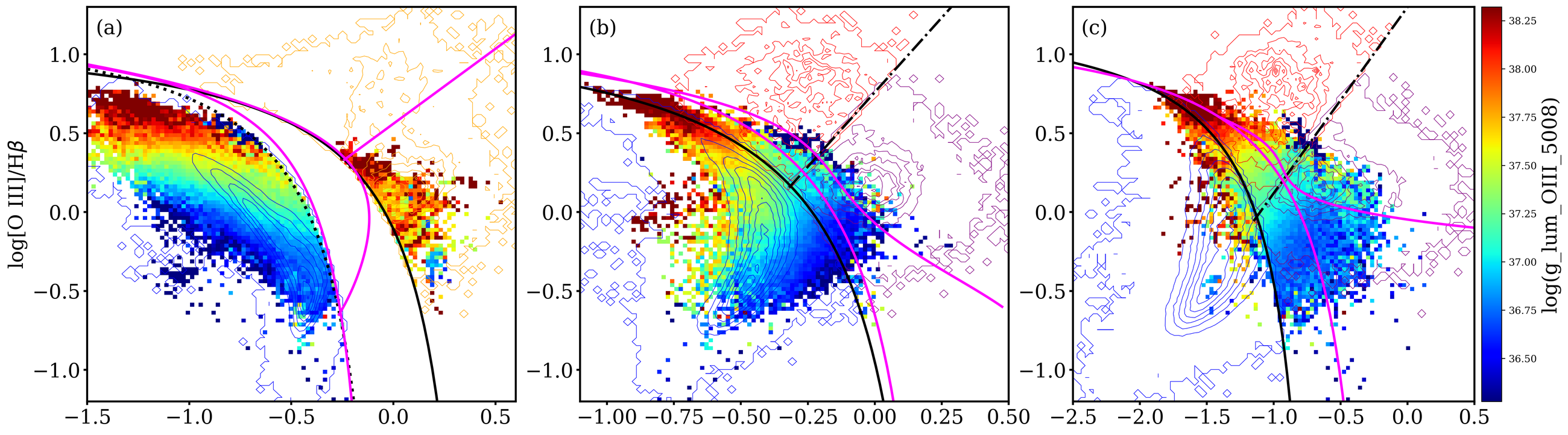}
\plotone{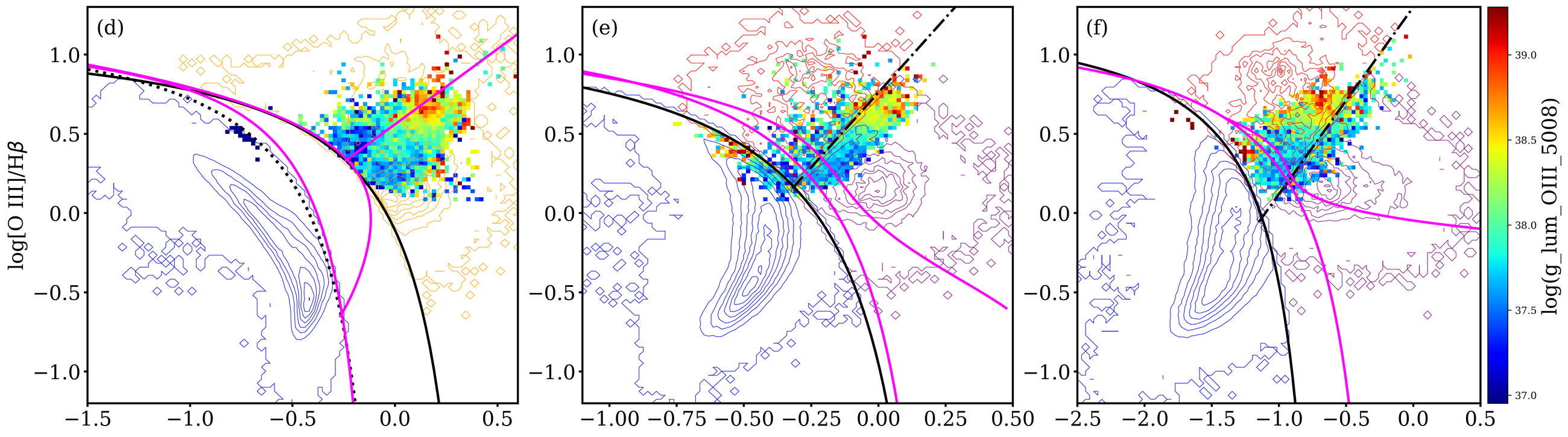}
\plotone{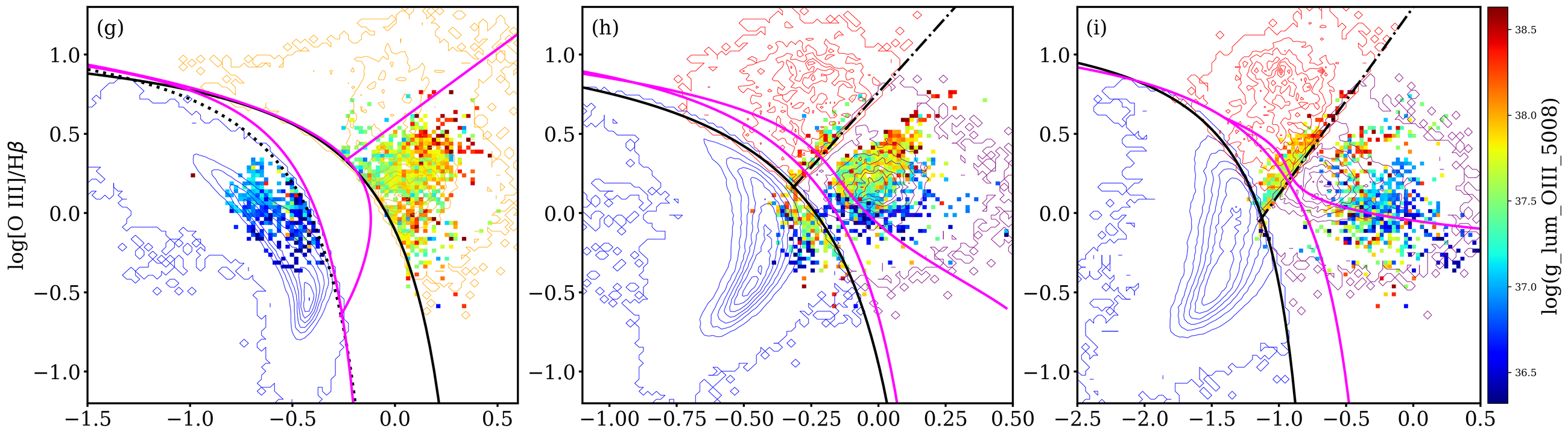}
\plotone{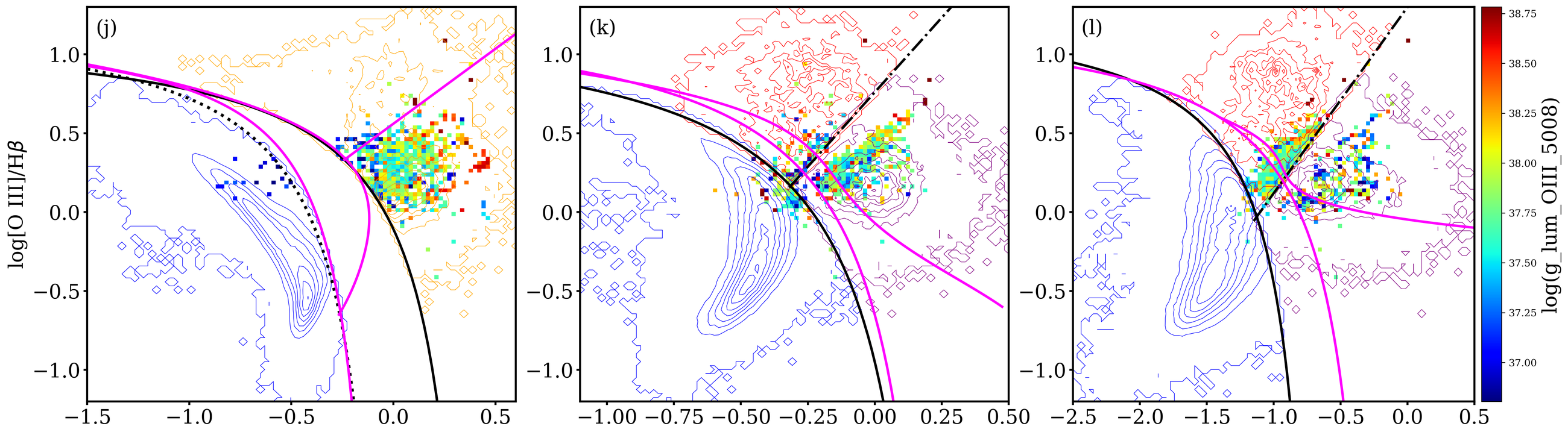}
\plotone{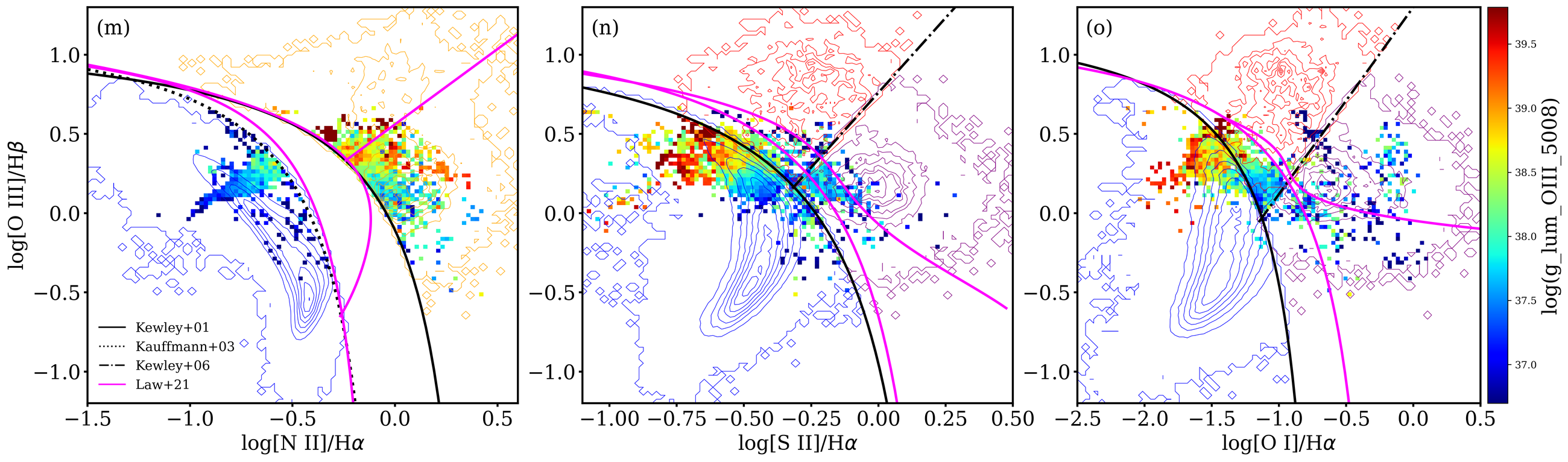}
\caption{Color-coded plots showing the intrinsic luminosity of the [O III] emission line for the five UMAP-classified categories: SF$^U$ (top row), Sy$^U$ (second row), LI(N)ER$^U$ (third row), U1 (fourth row), and U2 (bottom row).}
\label{fig:umap_amb_bpt_o3}
\end{figure*}

Figure \ref{fig:umap_amb_bpt_o3} presents color-coded plots based on the intrinsic luminosity of the [O III] emission line the SF$^U$, Sy$^U$, LI(N)ER$^U$, U1, and U2 clusters The [O III] luminosity serves as an important diagnostic for energetic ionization processes, often associated with AGN activity due to its high ionization potential, but it can also be elevated in low-metallicity SF regions. Thus, high [O III] luminosity can arise from distinct physical scenarios---either powerful AGN-driven photoionization \citep[e.g.,][]{kauffm2003, LaMassa2010} or intense, metal-poor star-formation \citep[e.g.,][]{Arata2020}.

For the SF$^U$ cluster (top row), the [N II]-BPT diagram shows a strong color gradient along the [O III]/H$\beta$ axis: spaxels with higher [O III]/H$\beta$ ratios tend to exhibit stronger [O III] luminosity. This trend reinforces the dual interpretation that high [O III] emission can stem from both low-metallicity SF regions and from AGN contamination near the boundary. The spaxels on the SF side show a broad spread in luminosity, whereas those extending into the AGN side are more consistently high-luminosity, likely reflecting AGN activity; however, these spaxels are not clearly identified in the [S II]- and [O I]-BPT diagrams.

For the Sy$^U$ cluster (second row), most spaxels lie within the AGN-designated regions of all three diagrams and tend to exhibit brighter [O III] luminosity, with a comparatively narrower color spread than in SF$^U$. A possible gradient is also visible, extending away from the SF demarcation lines.

In contrast, the LI(N)ER$^U$ spaxels (third row) display a more nuanced behavior. On the [N II]-BPT, the population appears bifurcated: one group shows relatively lower [O III] luminosities and occupies the SF-side (tentatively associated with LIERs), while another group on the AGN-side (potentially LINERs) shows moderately elevated [O III] luminosities. This dichotomy is consistent with what was observed in Figures \ref{fig:umap_amb_bpt_r} and \ref{fig:umap_amb_bpt_ha}, reinforcing the utility of the [N II]-BPT for distinguishing between LIER and LINER populations. The distinction becomes blurred in the [S II]-BPT diagram, where both groups overlap. However, in the [O I]-BPT (panel i), LINERs appear more concentrated in regions of elevated [O I]/H$\alpha$, offering tentative support for a possible separation in this plane as well.

For the U1 and U2 clusters (bottom two rows), no strong or consistent [O III] luminosity trends are evident. U1 (fourth row) appears more uniformly distributed with overall lower [O III] luminosities, suggesting weaker ionization conditions or a more heterogeneous physical origin. In contrast, U2 (bottom row) shows a mild [O III] luminosity gradient across all three BPT diagrams, particularly resembling the trend seen in the LI(N)ER$^U$ cluster. This behavior supports the interpretation that U2 consists of transitioning spaxels straddling the boundary between SF and AGN-dominated ionization regions.

\section{Discussions}
\label{sec:disc}

In Section \ref{sec:results}, we presented the UMAP model results for ambiguous spaxels and their positions on traditional BPT diagrams, showing that UMAP can successfully classify regions that may be misclassified or overlooked by traditional methods. By incorporating additional observational parameters, such as normalized elliptical radius, velocity dispersions, and [O III] luminosity, we demonstrated that these factors further aid in classification. This method not only enhances classification but also allows for the study of individual spaxels with unique properties, providing deeper insights into their characteristics. The inclusion of additional properties offers a more comprehensive understanding of these spaxels and opens new opportunities for detailed case studies.

\subsection{Individual case studies and incorporating additional properties for the UMAP model}
\label{subsec:future}

Following our study in Section \ref{sec:results}, we identified two intriguing subsets that suggest further investigation for deeper insights into the newly classified spaxels.

\begin{figure*}
\plotone{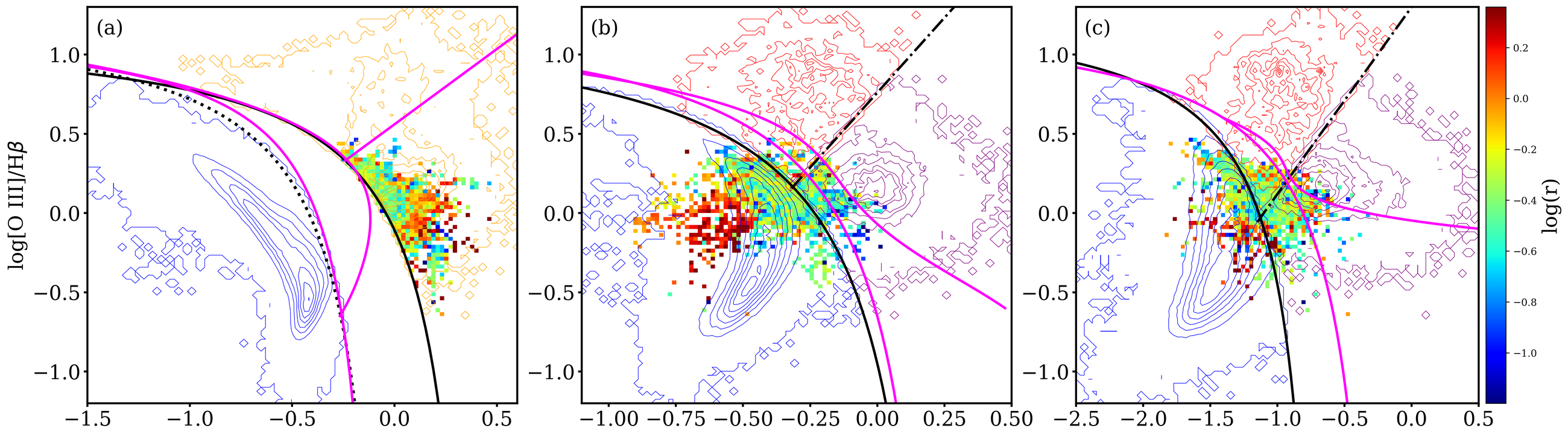}
\plotone{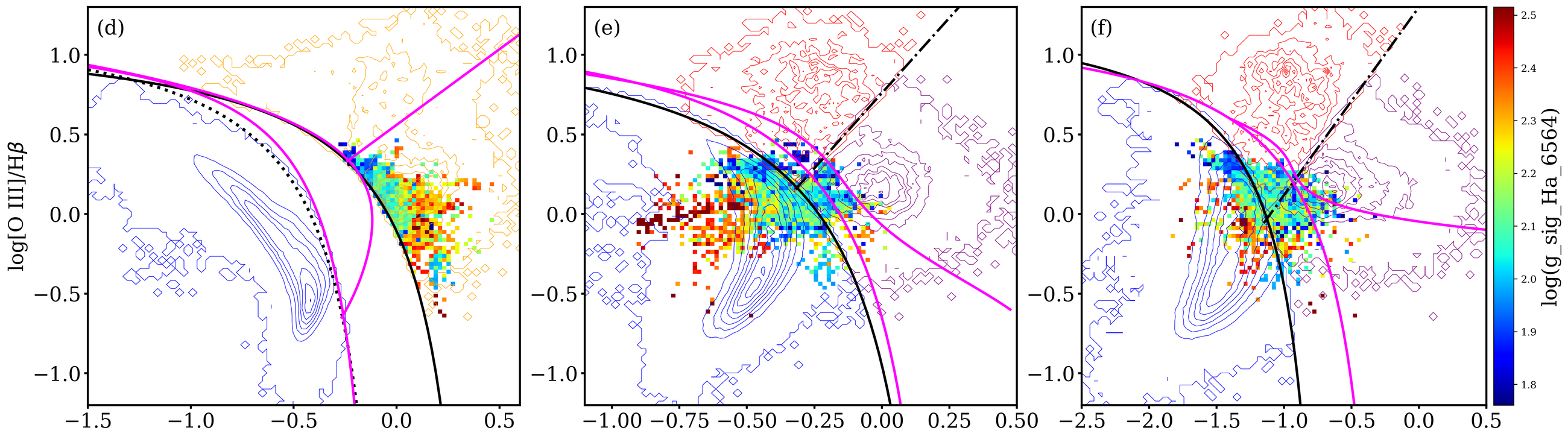}
\plotone{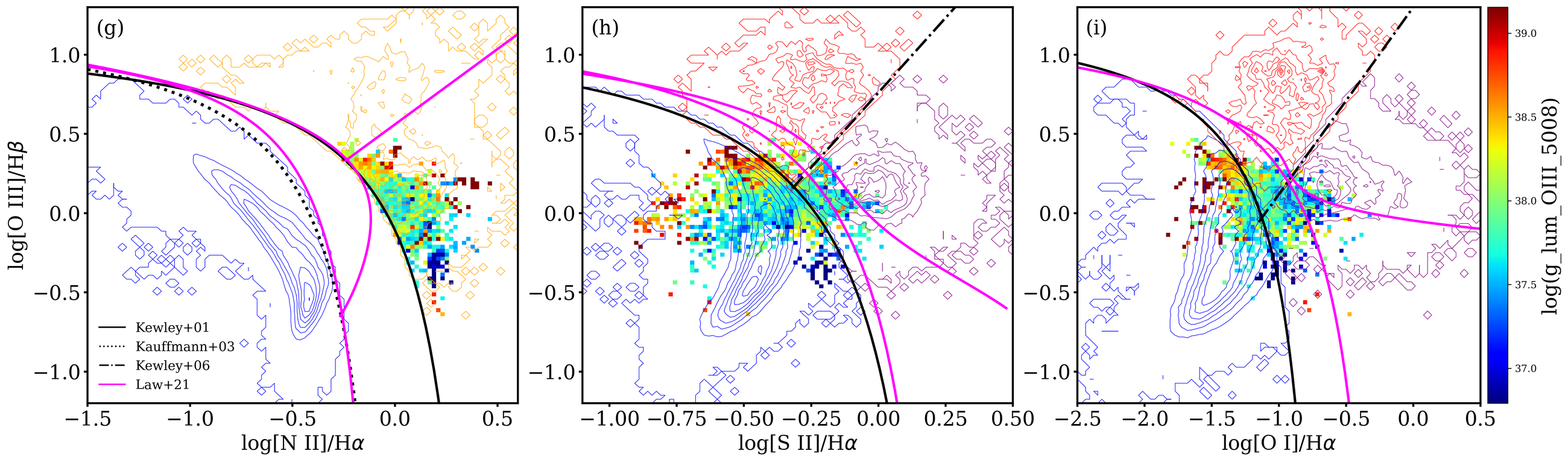}
\caption{Color-coded BPT diagrams showing the normalized elliptical radius (top), H$\alpha$ velocity dispersion (middle), and [O III] luminosity (bottom) for the SF$^U$ spaxels in the AGN.}
\label{fig:umap_amb_bpt_sf_sub1}
\end{figure*}

(i) SF$^U$ spaxels lying in the AGN region of the [N\,II]-BPT diagram:  Figure \ref{fig:umap_amb_bpt_sf_sub1} shows SF$^U$ spaxels that fall on the AGN side of the [N II]-BPT diagram while remaining within or near the SF locus on the [S II]- and [O I]-BPT diagrams, consistent with the well-known overlap of composite systems. Their H$\alpha$ velocity dispersions span $\sim$100–300 $\mathrm{km\,s^{-1}}$, and a coherent pattern emerges: broader lines are more common at larger radii ($>1\,r_e$) and tend to sit deeper within the SF region on the [S II]-BPT diagram, whereas spaxels closer to the center ($<1\,r_e$) typically show lower dispersions ($<100~\mathrm{km\,s^{-1}}$) and occupy buffer regions across all three diagrams. The [O III] luminosities lie in the range $10^{37}$–$10^{39}$ erg s$^{-1}$.

The combination of broad H$\alpha$ lines and locations away from the nucleus suggests that direct AGN influence is not required for these spaxels. Instead, the elevated dispersions are consistent with shock excitation or turbulence associated with interactions/mergers \citep[e.g.,][]{rich2011,Rich2014}. Indeed, a subset originates from the merging system with MaNGA ID \texttt{1-614567}; additional details are provided in Appendix \ref{app:individual}.

\begin{figure*}
\plotone{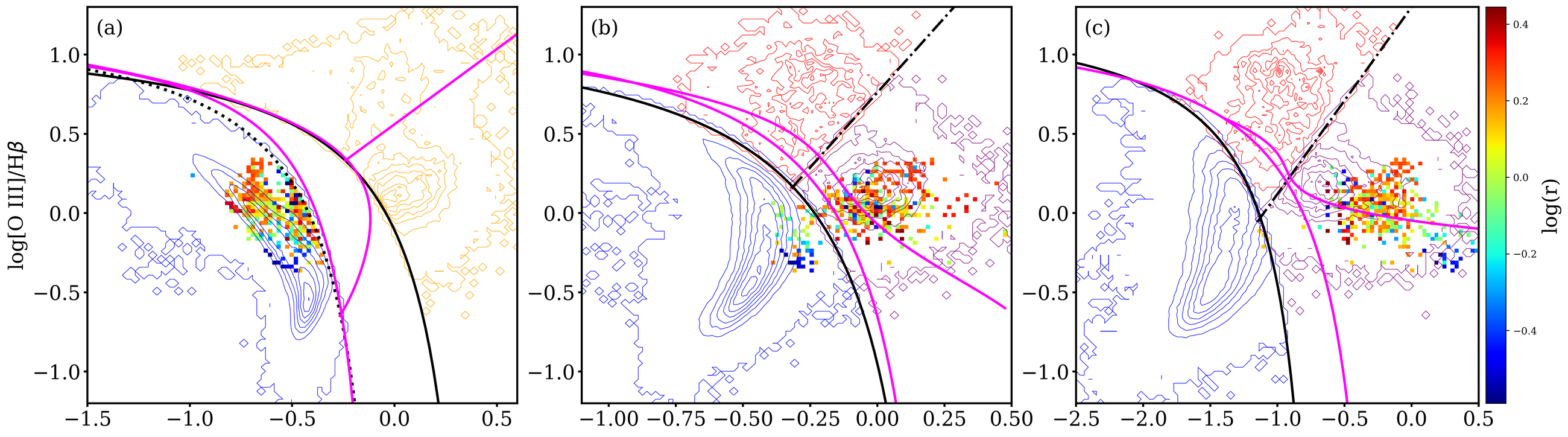}
\plotone{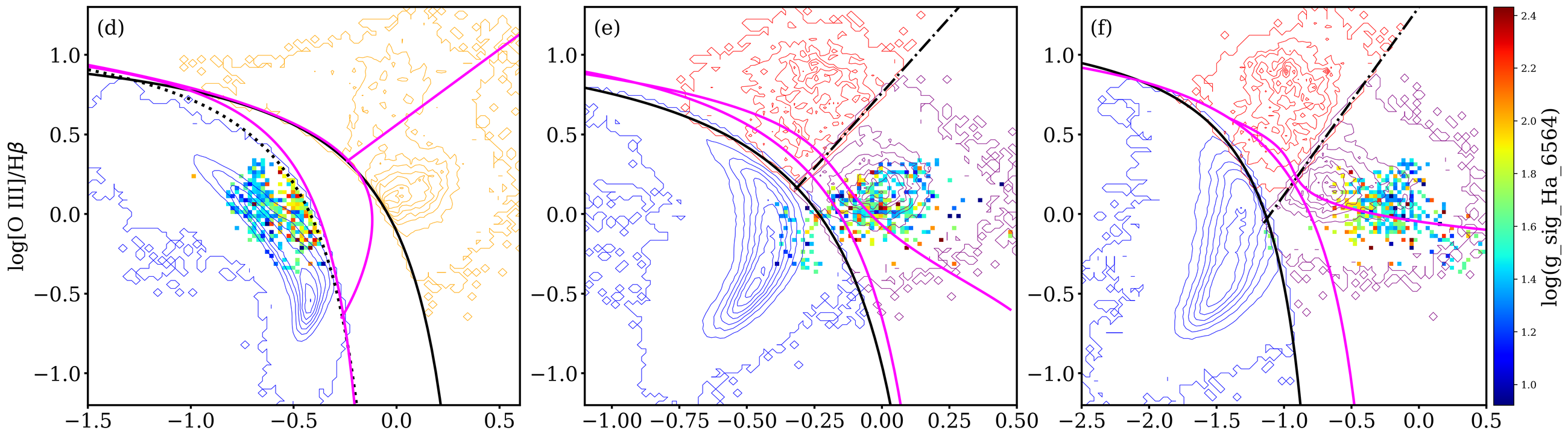}
\plotone{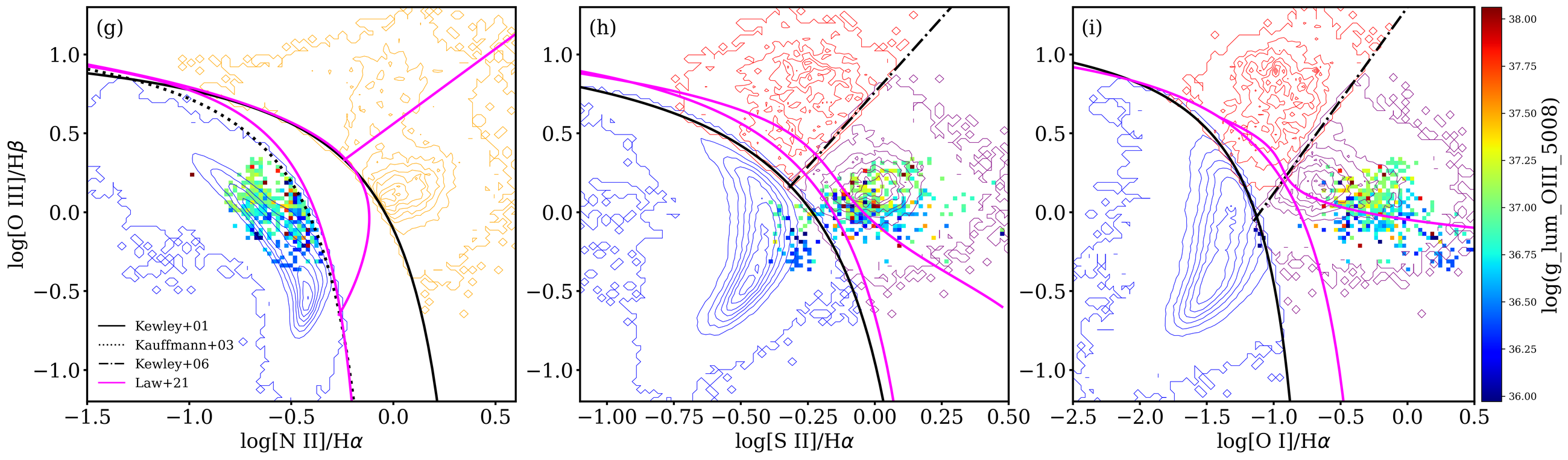}
\caption{Color-coded BPT diagrams showing the normalized elliptical radius (top), H$\alpha$ velocity dispersion (middle), and [O III] luminosity (bottom) for the LI(N)ER$^U$ spaxels in the SF.}
\label{fig:umap_amb_bpt_li_sub1}
\end{figure*}

(ii) LI(N)ER$^U$ spaxels classified as SF by the [N II]-BPT diagram: Figure \ref{fig:umap_amb_bpt_li_sub1} presents the LI(N)ER$^U$ subset that the [N II]-BPT diagram places on the SF side. These spaxels are predominantly found at large galactocentric distances ($\sim$1–3~$r_e$), exhibit narrow H$\alpha$ lines (typically 10–50 $\mathrm{km\,s^{-1}}$), and have [O III] luminosities of $10^{36}$–$10^{37}$~erg~s$^{-1}$. Their extended radii and narrow line widths argue against outflows or strong shocks, and instead align with ionization by hot  low-mass evolved stellar populations (HOLMES), as commonly inferred for extended LIER emission \citep{cidfernandes2011, Belfiore2016}. Thus, these spaxels are more naturally interpreted as LIERs rather than nuclear LINERs.

These case studies indicate that parameters such as velocity dispersion, galactocentric radius, and [O III] luminosity add informative context to BPT positions and can be incorporated into the dimensional-reduction step. Additional diagnostics may also be valuable; for example, the [O III]/[O II] ratio has been used to separate ionization states on sub-kpc scales \citep[e.g.,][]{kraeme2008}. Expanding the feature set and retraining/testing UMAP with these properties will be the focus of future work.

\subsection{Composite spaxels} \label{sec:comp}

While a detailed investigation of composite spaxels will be presented in a forthcoming paper, we have also applied the same UMAP-trained model to the composite spaxels to explore their behavior in the UMAP space.

\begin{figure}
    \centering
    \includegraphics[width=0.45\textwidth]{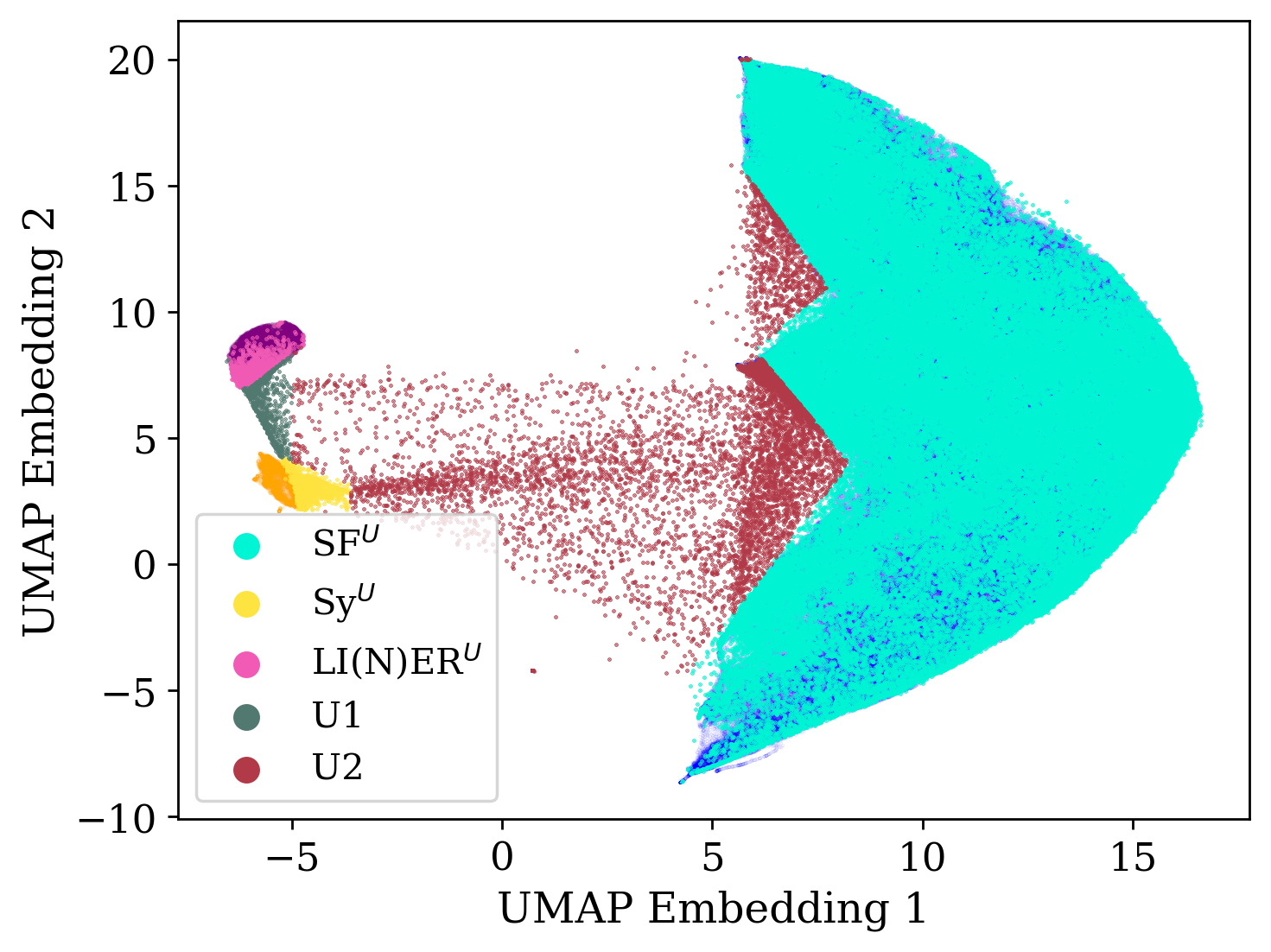}
    \caption{UMAP embeddings for composite spaxels, analogous to those shown in Figure \ref{fig:umap_combined}.}
    \label{fig:umap_comp}
\end{figure}

Figure \ref{fig:umap_comp} shows the UMAP embeddings for composite spaxels. In this figure, blue, orange, and purple dots represent SF, Sy, and LI(N)ER spaxels, respectively, with the testing set (composite spaxels) overlaid on the training set—analogous to the format shown in Figure \ref{fig:umap_combined}.

To extend the analysis, we assigned classifications to the overlaid composite spaxels based on their locations within the UMAP space. As in Figure \ref{fig:umap_combined}, the turquoise (SF$^U$), gold (Sy$^U$), and pink (LI(N)ER$^U$) dots indicate composite spaxels that fall within the previously defined SF, Sy, and LI(N)ER regions, respectively. These spaxels are newly categorized into these classes, offering distinctions that the traditional BPT diagrams could not resolve.

A subset of spaxels remains outside the previously defined UMAP clusters. To examine their characteristics more closely, we divided these unclassified spaxels into two subgroups: U1 (dark green) and U2 (brown). U1 spaxels tend to reside near the AGN region, whereas U2 spaxels are located in the intermediate space between AGN and SF regions.

Table \ref{tab:comp} summarizes the number of composite spaxels that are newly classified using the UMAP-based approach. Table \ref{tab:comp} reveals that the fraction of unclassified composite spaxels is nearly seven times higher than in the ambiguous spaxel case (Table \ref{tab:ambi}), indicating that composite spaxels are more difficult to classify within the UMAP framework. This supports the idea that composite spaxels occupy more transitional or overlapping regions in diagnostic space.

\begin{figure*}
\centering
\includegraphics[width=0.75\textwidth]{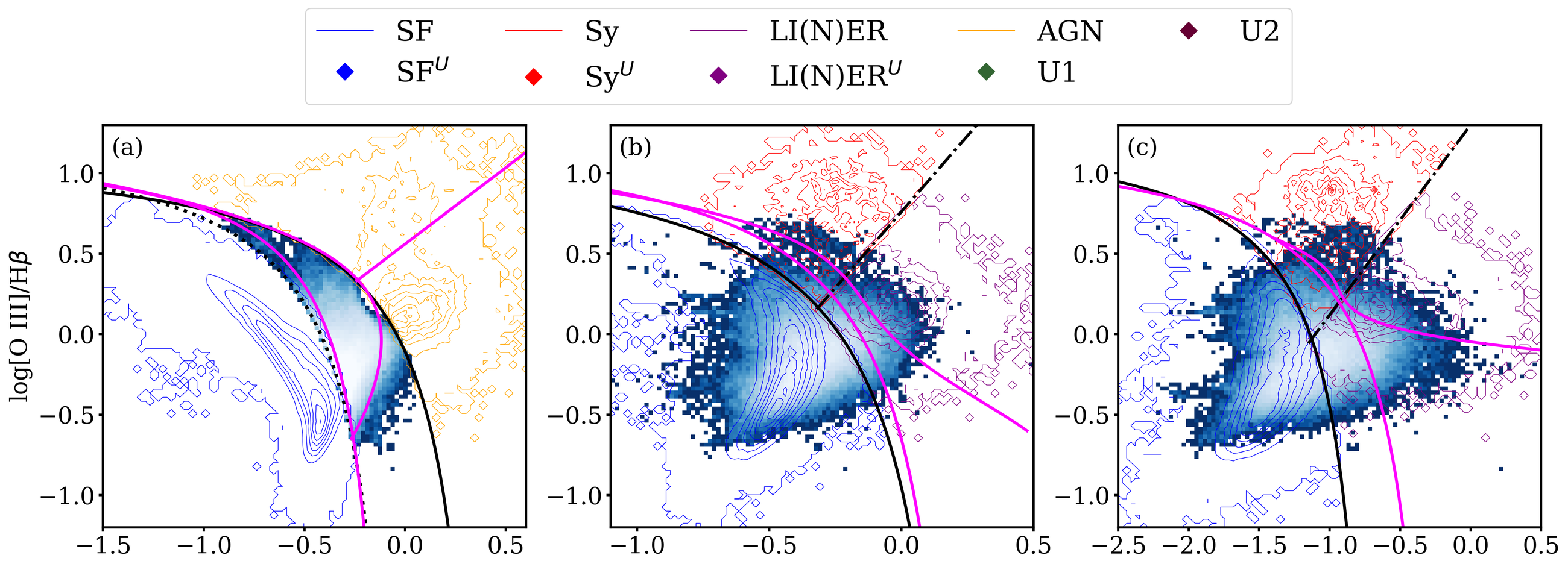}
\includegraphics[width=0.75\textwidth]{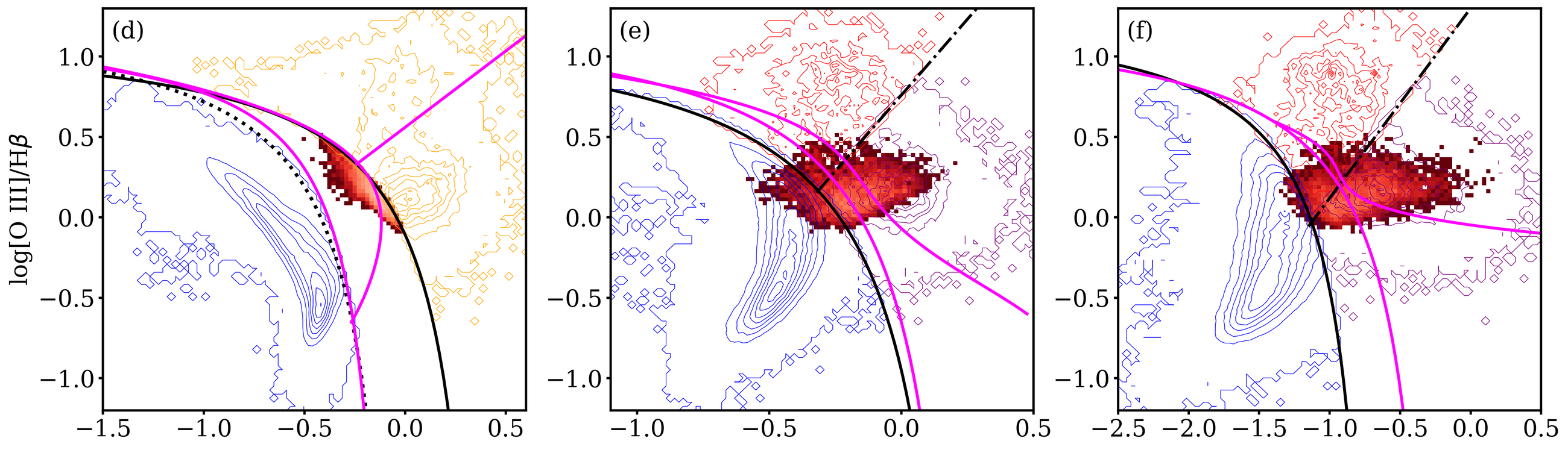}
\includegraphics[width=0.75\textwidth]{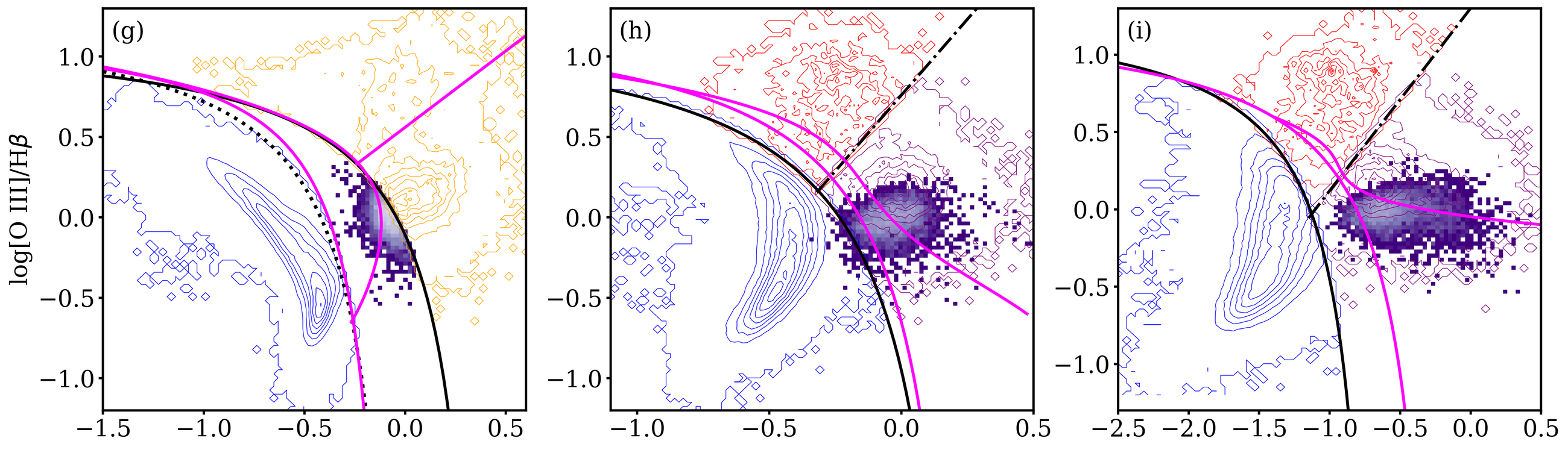}
\includegraphics[width=0.75\textwidth]{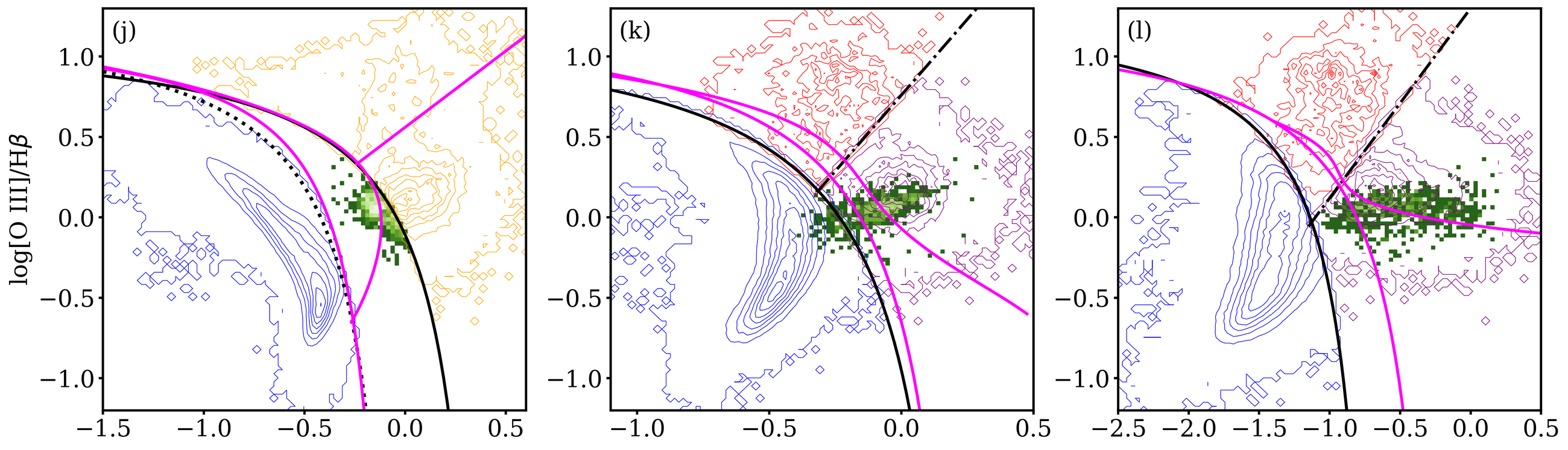}
\includegraphics[width=0.75\textwidth]{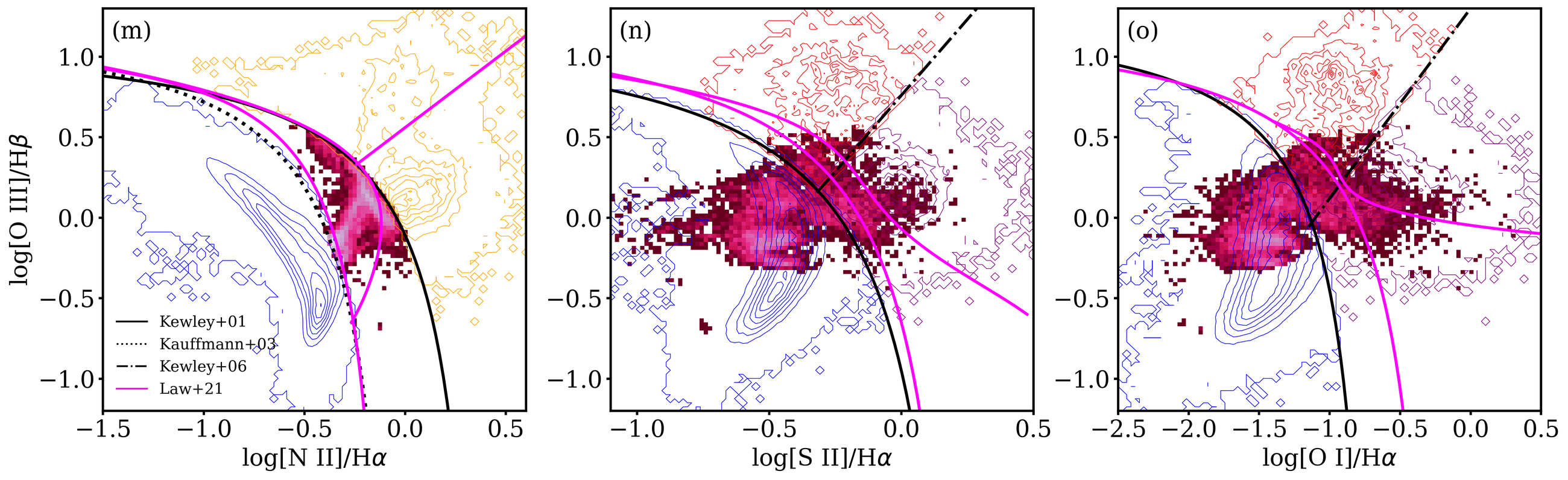}
\caption{The three BPT diagrams illustrating the distribution of composite spaxels reclassified by UMAP into SF$^U$ (first row), Sy$^U$ (second row), and LI(N)ER$^U$ (third row), as well as composite spaxels that remain unclassified, designated as U1 (fourth row) and U2 (fifth row).}
\label{fig:umap_comp_bpt}
\end{figure*}

To examine the distribution of spaxels originally classified as composite but newly reclassified by the UMAP method, we plotted their positions on the BPT diagrams using 2D binning (Figure \ref{fig:umap_comp_bpt}). Similar to Figure \ref{fig:umap_amb_bpt}, the classifications identified by UMAP---SF$^U$ (blue dots), Sy$^U$ (red dots), and LI(N)ER$^U$ (purple dots)---are shown, along with two subgroups of unclassified spaxels: U1 (green dots) and U2 (magenta dots). The underlying 2D histograms use an inverted color scheme, with white indicating regions of highest spaxel density. Contours representing the traditionally classified clean dataset are overlaid for direct comparison, visually highlighting the overlap and differences between traditional and UMAP-based classifications.

As with the ambiguous case, the UMAP-classified categories SF$^U$, Sy$^U$, and LI(N)ER$^U$ are not strictly confined within the established boundaries defined by \citet{kewley2001}, \citet{kewley2006} and \citet{kauffm2003}. Across all BPT diagrams, we observe instances where the established demarcation lines are crossed, and spaxels appear in regions typically assigned to other classes. This discrepancy likely results from the inherently composite nature of these spaxels. Nonetheless, their successful classification by the UMAP algorithm indicates that traditional diagnostic curves may require refinement or more nuanced interpretations to accurately categorize these cases. Compared to the ambiguous spaxels, these composite spaxels exhibit more compact distributions, possibly due to their inherent clustering within the composite region of the [N II]-BPT diagram. This highlights the challenges faced when classifying composite spaxels using conventional demarcation schemes and emphasizes the limitations of previous approaches. In a forthcoming paper, we will further investigate these composite spaxels by incorporating additional galaxy properties into the UMAP training set to improve classification.

\begin{deluxetable}{lc}
\tablenum{6}
\tablecaption{Number of composite samples classified by the model during analysis}
\label{tab:comp}
\tablewidth{0pt}
\tablehead{
{Name} & {Number}}
\startdata
SF$^U$  & 163 424 (94.4\%) \\
Sy$^U$ & 4 846 (2.80\%) \\
LI(N)ER$^U$ & 1 789 (1.03\%) \\
Unclassified 1 (U1)& 876 (0.51\%) \\
Unclassified 2 (U2)& 2 194 (1.27\%) \\
\hline
Total & 173 129 (100\%) \\
\enddata
\end{deluxetable}

\section{Conclusion}
\label{sec:conclusion}

In this study, we investigate the use of unsupervised machine learning to improve the classification of optical spectra and gain deeper insights into ionization regions within galaxies. Our aim is to present a complementary approach that addresses the limitations of traditional BPT diagnostics, which often struggle with boundary inconsistencies caused by overlapping ionization effects.

We used only clean data---spaxels consistently classified as SF, Sy, or LI(N)ER across all three BPT diagrams---to train the UMAP model. This approach aimed to enhance the model’s reliability, as including composite and ambiguous data could introduce contamination, blurring the boundaries between classes and complicating clear classification. By excluding such uncertain data, we reduced noise, achieving a clearer and more robust classification outcome (upper panel of Figure \ref{fig:umap_combined}).

After training the UMAP model exclusively with clean data, we tested it on ambiguous data. Remarkably, many of these ambiguous spaxels were classified into well-defined regions (lower panel of Figure \ref{fig:umap_combined}), rather than remaining in the uncertain spaces between classes. This suggests that the model is effective in classifying previously ambiguous spaxels.%

Testing the ambiguous spaxels further underscores the inconsistencies of the traditional BPT-based classification method. In several cases, UMAP assigned classifications to spaxels that conflicted with the boundaries established by previous studies, underscoring the limitations of relying solely on pre-defined BPT curves and highlighting the potential of data-driven approaches to reveal inconsistencies in these diagnostics. Consequently, we introduce new boundary equations to provide a more accurate classification framework, offering guidance for readers seeking to classify their data (see Appendix \ref{app:boundary}).

An additional finding of this study is the emergence of a possible dichotomy within the LI(N)ER$^U$ population. As shown in the third row of Figure \ref{fig:umap_amb_bpt_r}, spaxels classified as LI(N)ER$^U$ appear to separate into two groups on the [N II]-BPT diagram: one group lies closer to the SF side of the demarcation line, while the other tends to occupy the AGN region. While this separation may reflect a distinction between LIERs and LINERs, there is considerable overlap, and some LIERs appear on the AGN side. This trend is further supported by other diagnostics: in Figure \ref{fig:umap_amb_bpt_ha}, spaxels likely corresponding to LIERs show narrower H$\alpha$ line widths, whereas those likely corresponding to LINERs exhibit broader profiles. In addition, Figure \ref{fig:umap_amb_bpt_o3} reveals that LIERs tend to have lower [OIII] luminosities and are located at larger galactocentric radii (higher $\log(r)$) compared to LINERs. While overlap persists on the [S II]-BPT diagram, the [O I]/H$\alpha$ ratios show a tentative trend with LIERs occupying slightly elevated regions. We note, however, that LIER emission can also extend onto the AGN side of the [N II]-BPT diagram. These results suggest that the [N II]-BPT, when combined with kinematic and spatial information, may provide a promising axis for distinguishing between different low-ionization sources. Future studies incorporating multi-dimensional diagnostics---such as [O I]/H$\alpha$ and effective radius---could offer a clearer physical interpretation of the LI(N)ER population and further refine their classification. If the model can accurately distinguish these subtle differences and correctly identify the underlying physics, it could offer a deeper understanding of each ionization region within a galaxy. This approach could supplement the traditional BPT diagrams, providing a clearer and more effective method for analyzing galaxy ionization regions and opening new opportunities for research.

The new method enables the identification of peculiar behaviors in individual spaxels, revealing insights that the traditional BPT diagram scheme might have missed. By integrating additional physical parameters, we identified SF$^U$ spaxels with broad H$\alpha$ emission lines situated farther from the galactic center. Further investigation revealed that these spaxels originated from a galaxy undergoing a merger (see Appendix \ref{app:individual}), demonstrating the method’s effectiveness in uncovering complex behaviors that were not evident with BPT classification alone.

UMAP’s flexibility supports the inclusion of new parameters, enabling further refinement and scalability. Future work will aim to incorporate additional variables, such as other emission line ratios (e.g., [O II]/[O III]) or kinetic parameters (e.g., velocity dispersions) to enhance model training. It is essential to select these parameters based on established physical principles to avoid complicating the interpretation of the underlying physics. The scalability of machine learning methods is a key advantage, as they can deliver more accurate classifications with increased observational data. However, careful sample selection remains crucial to prevent overfitting and maintain the model’s ability to new data.

More broadly, our findings reveal that inconsistencies between BPT diagrams are not fully resolved by traditional or even extended classification schemes such as the 3D diagnostic method by \citet{law2021_refine}, which incorporates kinematic information. While such methods represent an important advancement, they still leave significant ambiguities near boundary regions. Our UMAP-based approach, in contrast, provides a data-driven alternative that captures non-linear relationships across multiple diagnostics, without requiring predefined boundaries. This reinforces the value of dimensionality reduction techniques in re-evaluating long-standing classification schemes, and it motivates future efforts to integrate both emission line ratios and additional physical parameters into unified, interpretable models of ionization structure.

\begin{acknowledgments}
    We thank the anonymous referee for their thoughtful comments and constructive suggestions, which helped improve the clarity and quality of this manuscript.
\end{acknowledgments}

\vspace{5mm}

\software{ \texttt{astropy} \citep{astropy}, \texttt{numpy} \citep{numpy}, \texttt{pandas} \citep{reback2020pandas}, \texttt{umap-learn} \citep{umap}, \texttt{hdbscan} \citep{hdbscan}, \texttt{scipy} \citep{2020SciPy-NMeth}
}

\appendix

\section{Defining Boundaries for Newly Identified Uncertain Regions}
\label{app:boundary}

In this appendix, we outline the methodology for deriving new buffer equations from the UMAP results. We provide a detailed explanation of how the UMAP projections are utilized to identify new buffer zones and how these zones inform the subsequent formulation of the buffer equations.

We begin by examining the UMAP classification results. Typically, a significant portion of the UMAP classifications aligns closely with traditional methods defined by \citet{kewley2001}, \citet{kewley2006}, and \citet{kauffm2003} on the BPT diagrams. However, there are notable instances where UMAP classifications deviate from these traditional diagnostic boundaries. To explore these discrepancies, we specifically focus on spaxels located outside the established boundaries, selecting those whose UMAP classifications differ from the conventional regions for further investigation.

Next, for spaxels classified into categories that do not align with the traditional BPT demarcation boundaries, we define new regions and derive corresponding buffer equations to quantify these boundaries. To achieve this, we measure the distances from each spaxel to the established demarcation lines on the BPT diagrams. We then select the closest 95 percent and, for comparison, 99 percent of these points to exclude outliers, ensuring that the defined buffer regions are robust and reliable.

Then, we calculate the convex hull of a set of filtered spaxels. The convex hull is the smallest convex boundary that encloses a set of points, forming the "outer shell" of the data by connecting the outermost points. This ensures that any line segment between two points within the hull lies entirely inside the boundary. The convex hull provides a useful boundary for clustering and classification, as it represents the outer limits of the data distribution and helps define regions of interest, identify extreme points, and efficiently process spatial relationships. The convex hull algorithm used in this analysis is implemented via the \texttt{Python} package \texttt{scipy} \citep{2020SciPy-NMeth}.

Lastly, using the vertices obtained from the convex hull, we utilize the boundary points to derive the boundary fitting equations. While we aim to keep the equations as simple as possible, representing most boundaries with second- to fourth-order polynomials, higher-order (up to sixth-order) fits are used when required for accuracy. Some vertices are excluded from this process as they lay on pre-existing boundaries defined by \cite{kewley2001}, \cite{kewley2006} and \cite{kauffm2003}, and thus are not useful for generating new boundary fitting equations.

\begin{figure*}
\plotone{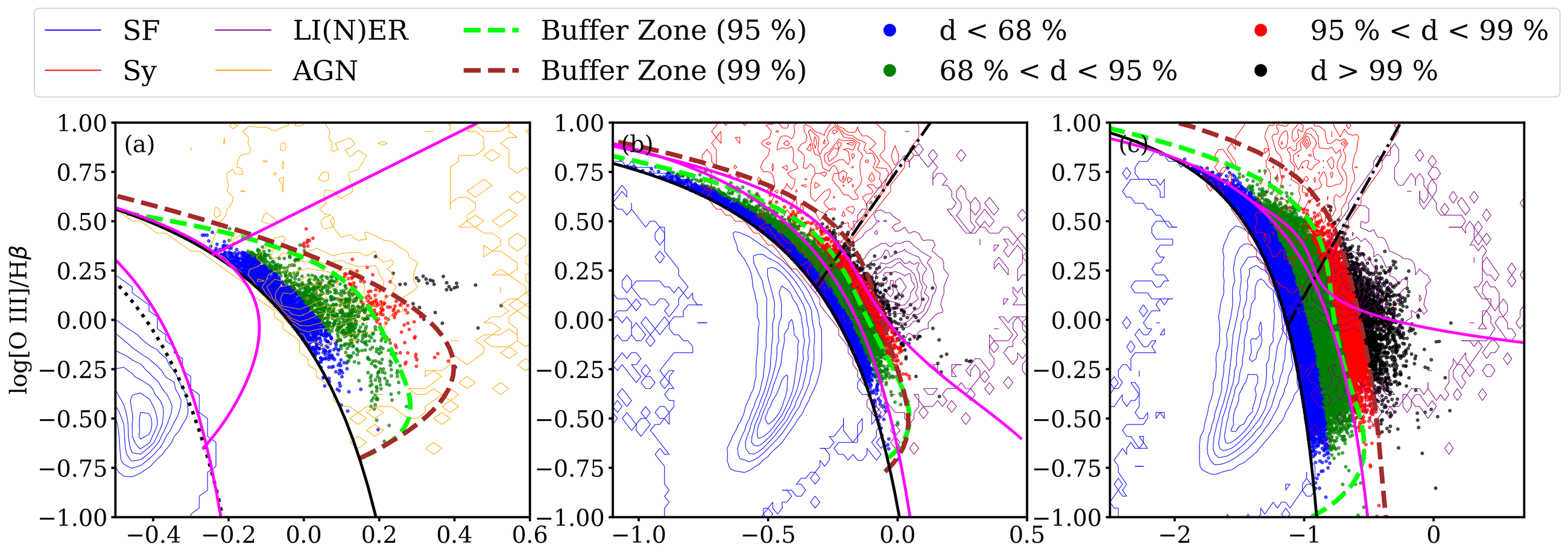}
\plotone{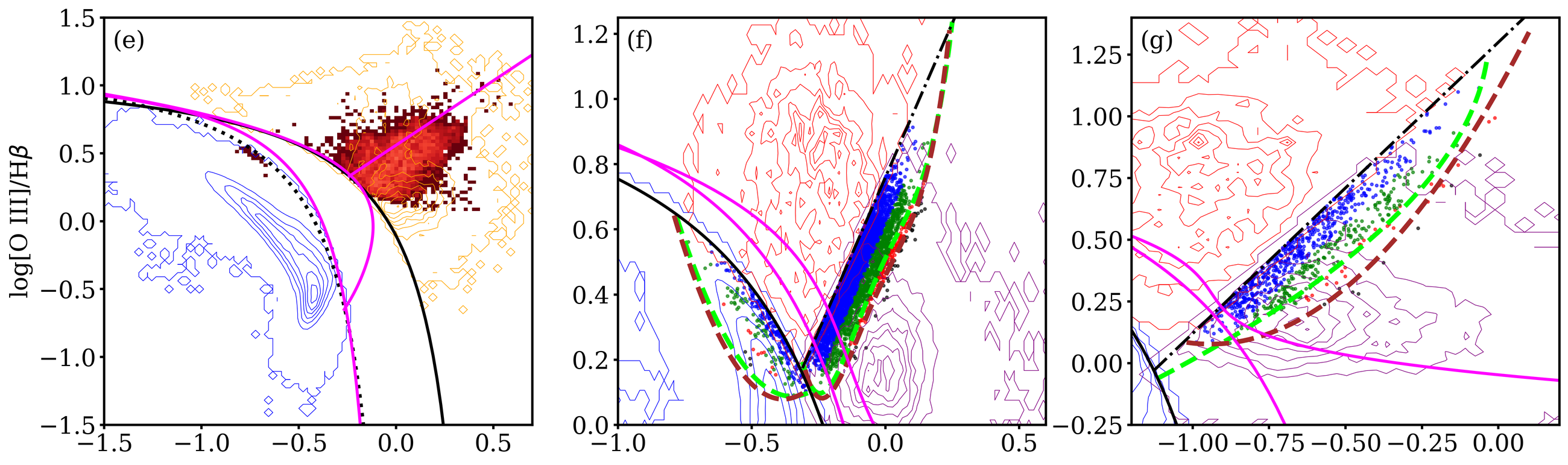}
\plotone{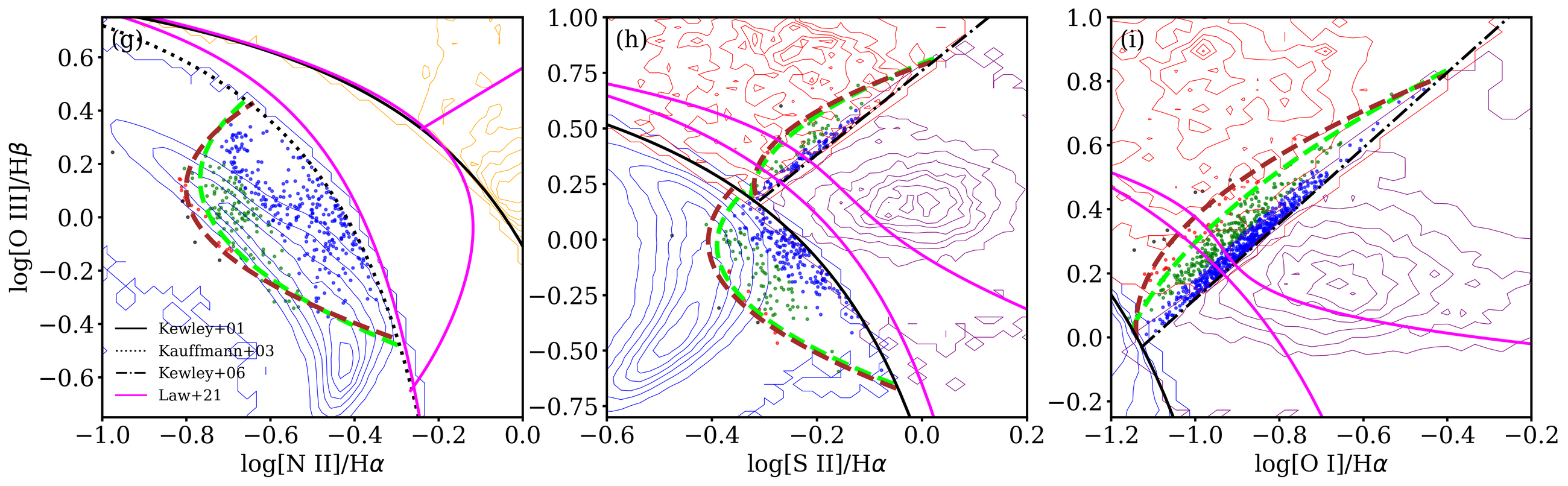}
\caption{Zoomed-in views of the three BPT diagrams showing the distribution of ambiguous spaxels classified by UMAP into SF$^U$ (first row), Sy$^U$ (second row), and LI(N)ER$^U$ (third row). Blue, green, and red dots indicate spaxels within the 68\%, 95\%, and 99\% percentiles of proximity to the predefined boundaries, respectively, while black dots mark the outermost 1\% of spaxels. The lime and brown dashed curves represent the UMAP-derived buffer equations enclosing 95\% and 99\% of the spaxels. For the Sy$^U$ [N II]-BPT diagram, no new buffer curves are defined because the spaxels do not extend sufficiently beyond the AGN region; these spaxels are still shown for clarity and completeness.}
\label{fig:umap_amb_zoom}
\end{figure*}

Here, we present the new buffer equations derived from the procedure described above, as shown in Figure \ref{fig:umap_amb_bpt}. To examine the spatial distribution in greater detail, Figure \ref{fig:umap_amb_zoom} provides a zoomed-in view highlighting how the spaxels are arranged relative to the boundaries. In this figure, blue dots represent the 68\% of spaxels closest to the predefined boundaries, green dots correspond to those between the 68th and 95th percentiles, and red dots mark spaxels between the 95th and 99th percentiles. The remaining 1\% of spaxels, which lie furthest from the demarcation curves, are shown in black. As in Figure \ref{fig:umap_amb_bpt}, the lime and brown dashed lines in Figure \ref{fig:umap_amb_zoom} also indicate the UMAP-derived buffer equations enclosing 95\% and 99\% of the spaxels, respectively. However, note that the Sy$^U$ spaxels on the [N II]-BPT diagram do not extend sufficiently beyond the AGN-classified region to define buffer curves. Nevertheless, we include the Sy$^U$ spaxels in the figure for clarity and completeness.

\subsection{Buffer equations enclosing 95\% of spaxels}

For SF$^{U}$ but not in the SF regions on the BPT diagrams:

\begin{eqnarray}
    \log(\text{[N II]/H}\alpha)  < & -2.2 [\log(\text{[O III]/H}\beta)]^4 - 1.3 [\log(\text{[O III]/H}\beta)]^3 - 0.36 [\log(\text{[O III]/H}\beta)]^2 \\ & - 0.29 \log(\text{[O III]/H}\beta) + 0.19,  \quad \text{for } -0.70 < \log(\text{[O III]/H}\beta) < 0.54. \nonumber
\end{eqnarray}

\begin{eqnarray}
    \log(\text{[S II]/H}\alpha) < & -1.3 [\log(\text{[O III]/H}\beta)]^4 - 0.24 [\log(\text{[O III]/H}\beta)]^3 + 0.11 [\log(\text{[O III]/H}\beta)]^2 \\& - 0.37 \log(\text{[O III]/H}\beta) - 0.11,  \quad
    \text{for } -0.70 < \log(\text{[O III]/H}\beta) < 0.90. \nonumber
\end{eqnarray}

\begin{eqnarray}
    \log(\text{[O I]/H}\alpha) < & -1.44 [\log(\text{[O III]/H}\beta)]^4 - 0.66 [\log(\text{[O III]/H}\beta)]^3 + 0.38 [\log(\text{[O III]/H}\beta)]^2 \\ & - 0.23 \log(\text{[O III]/H}\beta) - 0.77, \quad
    \text{for } -1.0 < \log(\text{[O III]/H}\beta) < 1.0. \nonumber
\end{eqnarray}

For Sy$^{U}$ but in the LI(N)ER regions on the BPT diagrams:

\begin{eqnarray}
    \log(\text{[O III]/H}\beta) > & 75.62 [\log(\text{[S II]/H}\alpha)]^4 + 9.10 [\log(\text{[S II]/H}\alpha)]^3 - 2.24 [\log(\text{[S II]/H}\alpha)]^2 \\ & + 1.71 \log(\text{[S II]/H}\alpha)  +0.51,
    \quad  \text{for } -0.31 < \log(\text{[S II]/H}\alpha) < 0.25. \nonumber
\end{eqnarray}

\begin{eqnarray}
    \log(\text{[O III]/H}\beta) > & \ 2.96 [\log(\text{[S II]/H}\alpha)]^2 + 2.07 \log(\text{[S II]/H}\alpha) +0.45, \\ &
     \text{for } -0.78 < \log(\text{[S II]/H}\alpha) < 0.29. \nonumber
\end{eqnarray}

\begin{eqnarray}
    \log(\text{[O I]/H}\alpha) < & -0.55 [\log(\text{[O III]/H}\beta)]^2 + 1.48 \log(\text{[O III]/H}\beta) - 1.02, \\
    & \text{for } -0.06 < \log(\text{[O III]/H}\beta) < 1.25. \nonumber
\end{eqnarray}

For LI(N)ER$^U$ but not in the LI(N)ER regions on the BPT diagrams:

\begin{eqnarray}
    \log(\text{[N II]/H}\alpha) > & \ 1.22 [\log(\text{[O III]/H}\beta)]^2  - 0.35 \log(\text{[O III]/H}\beta) -0.74,  \\  \nonumber
    & \text{for } -0.48 < \log(\text{[O III]/H}\beta) < 0.44. \nonumber
\end{eqnarray}

\begin{eqnarray}
    \log(\text{[S II]/H}\alpha) > & \ 0.87 [\log(\text{[O III]/H}\beta)]^2 + 0.05 \log(\text{[O III]/H}\beta) -0.39, \\
    & \text{for } -0.65 < \log(\text{[O III]/H}\beta) < 0.22. \nonumber
\end{eqnarray}

\begin{eqnarray}
    \log(\text{[S II]/H}\alpha) > & \ 0.83 [\log(\text{[O III]/H}\beta)]^2 - 0.28 \log(\text{[O III]/H}\beta) -0.30, \\
    & \text{for }  0.19 < \log(\text{[O III]/H}\beta) < 0.82. \nonumber
\end{eqnarray}

\begin{eqnarray}
    \log(\text{[O I]/H}\alpha) > & 0.69 [\log(\text{[O III]/H}\beta)]^2 +0.33 \log(\text{[O III]/H}\beta) - 1.16, \\
    & \text{for } 0.03 < \log(\text{[O III]/H}\beta) < 0.84. \nonumber
\end{eqnarray}

\subsection{Buffer equations enclosing 99\% of spaxels}

For SF$^{U}$ but not in the SF regions on the BPT diagrams:

\begin{eqnarray}
    \log(\text{[N II]/H}\alpha)  < & - 1.18 [\log(\text{[O III]/H}\beta)]^2  - 0.57 \log(\text{[O III]/H}\beta) + 0.32, \\ &\text{for } -0.70 < \log(\text{[O III]/H}\beta) < 1.00. \nonumber
\end{eqnarray}

\begin{eqnarray}
    \log(\text{[S II]/H}\alpha) < & -0.91 [\log(\text{[O III]/H}\beta)]^4 - 0.31 [\log(\text{[O III]/H}\beta)]^3 + 0.03 [\log(\text{[O III]/H}\beta)]^2 \\& - 0.23 \log(\text{[O III]/H}\beta) - 0.06,  \quad
    \text{for } -0.77 < \log(\text{[O III]/H}\beta) < 1.20. \nonumber
\end{eqnarray}

\begin{eqnarray}
    \log(\text{[O I]/H}\alpha) & < -0.22 [\log(\text{[O III]/H}\beta)]^6 - 0.51 [\log(\text{[O III]/H}\beta)]^5 - 0.27 [\log(\text{[O III]/H}\beta)]^4 \\ & + 0.04 [\log(\text{[O III]/H}\beta)]^3 - 0.11[\log(\text{[O III]/H}\beta)]^2 + 0.33 \log(\text{[O III]/H}\beta) - 0.57,  \nonumber \\ &
    \text{for } -1.70 < \log(\text{[O III]/H}\beta) < 1.50. \nonumber
\end{eqnarray}

For Sy$^{U}$ but in the LI(N)ER regions on the BPT diagrams:

\begin{eqnarray}
    \log(\text{[O III]/H}\beta) > & \ 3.40 [\log(\text{[S II]/H}\alpha)]^2 + 2.60 \log(\text{[S II]/H}\alpha) +0.58,
    \\ &  \text{for } -0.79 < \log(\text{[S II]/H}\alpha) < -0.29. \nonumber
\end{eqnarray}

\begin{eqnarray}
    \log(\text{[O III]/H}\beta) > & \ 81.57 [\log(\text{[S II]/H}\alpha)]^4 + 11.84 [\log(\text{[S II]/H}\alpha)]^3 -1.42 [\log(\text{[S II]/H}\alpha)]^2 \\ & + 1.67 \log(\text{[S II]/H}\alpha) +0.46,
     \quad \text{for } -0.30 < \log(\text{[S II]/H}\alpha) < 0.24. \nonumber
\end{eqnarray}

\begin{eqnarray}
    \log(\text{[O III]/H}\beta) > & \ 1.16 [\log(\text{[O I]/H}\alpha)]^2 + 2.19 \log(\text{[O I]/H}\alpha) + 1.11,
    \\ &  \text{for } -1.02 < \log(\text{[O I]/H}\alpha) < 0.10. \nonumber
\end{eqnarray}

For LI(N)ER$^U$ but not in the LI(N)ER regions on the BPT diagrams:

\begin{eqnarray}
    \log(\text{[N II]/H}\alpha) > & \ 1.57 [\log(\text{[O III]/H}\beta)]^2  - 0.34 \log(\text{[O III]/H}\beta) -0.78,  \\
    & \text{for } -0.45 < \log(\text{[O III]/H}\beta) < 0.43. \nonumber
\end{eqnarray}

\begin{eqnarray}
    \log(\text{[S II]/H}\alpha) > & \ 0.81 [\log(\text{[O III]/H}\beta)]^2 + 0.01 \log(\text{[O III]/H}\beta) -0.41, \\
    & \text{for } -0.67 < \log(\text{[O III]/H}\beta) < 0.23. \nonumber
\end{eqnarray}

\begin{eqnarray}
    \log(\text{[S II]/H}\alpha) > & \ 1.12 [\log(\text{[O III]/H}\beta)]^2 - 0.56 \log(\text{[O III]/H}\beta) -0.25, \\
    & \text{for }  0.20 < \log(\text{[O III]/H}\beta) < 0.80. \nonumber
\end{eqnarray}

\begin{eqnarray}
    \log(\text{[O I]/H}\alpha) > & 1.19 [\log(\text{[O III]/H}\beta)]^2 -0.08 \log(\text{[O III]/H}\beta) - 1.14, \\
    & \text{for } 0.00 < \log(\text{[O III]/H}\beta) < 0.82. \nonumber
\end{eqnarray}

\section{An Example of Applying This Diagnostic: Case Study of a Merging Galaxy}
\label{app:individual}

In Section \ref{subsec:future}, we explore the potential for studying specific spaxels exhibiting peculiar behavior using this diagnostic method. In particular, we focus on SF$^U$ spaxels located in the AGN of the [N II]-BPT diagram. As discussed earlier, a pattern emerges where broader H$\alpha$ emission lines correspond to spaxels positioned deeper into the SF of the [S II]-BPT diagram. A similar trend is observed with the spaxels' radii: the farther they are from the galaxy center, the deeper they appear within the SF of the [S II]-BPT diagram. Based on these observations, we selected spaxels that meet the following criteria: $\log \sigma_{\text{H}\alpha} > 2.6$. After filtering, 57 spaxels were selected; 56 of these spaxels are from the MaNGA ID \texttt{1-614567}, and the remaining one is from \texttt{1-43214}.

\begin{figure*}
\plotone{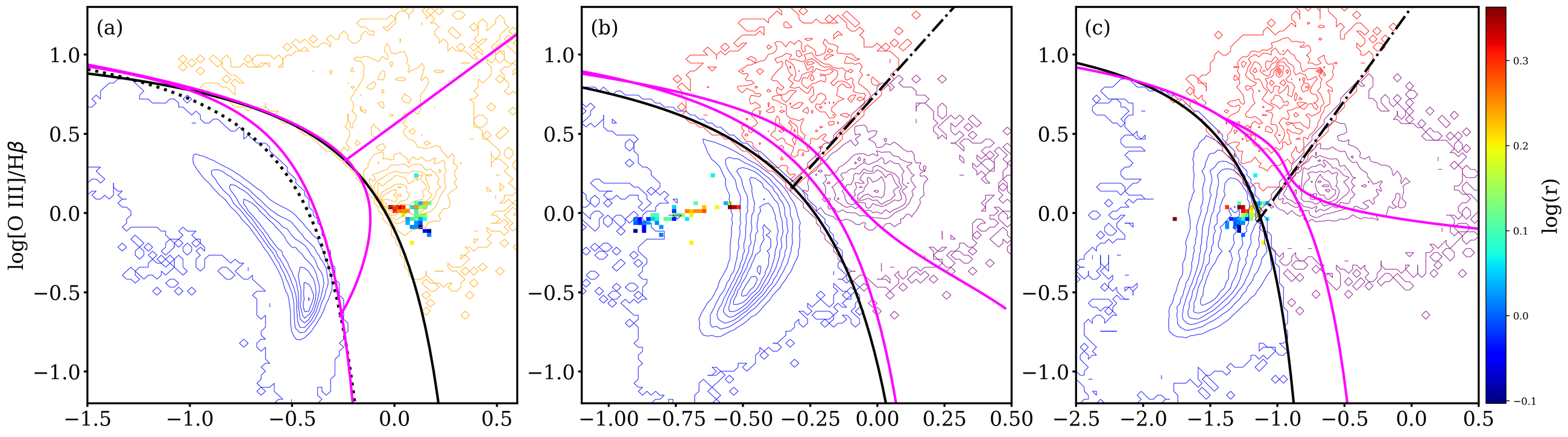}
\plotone{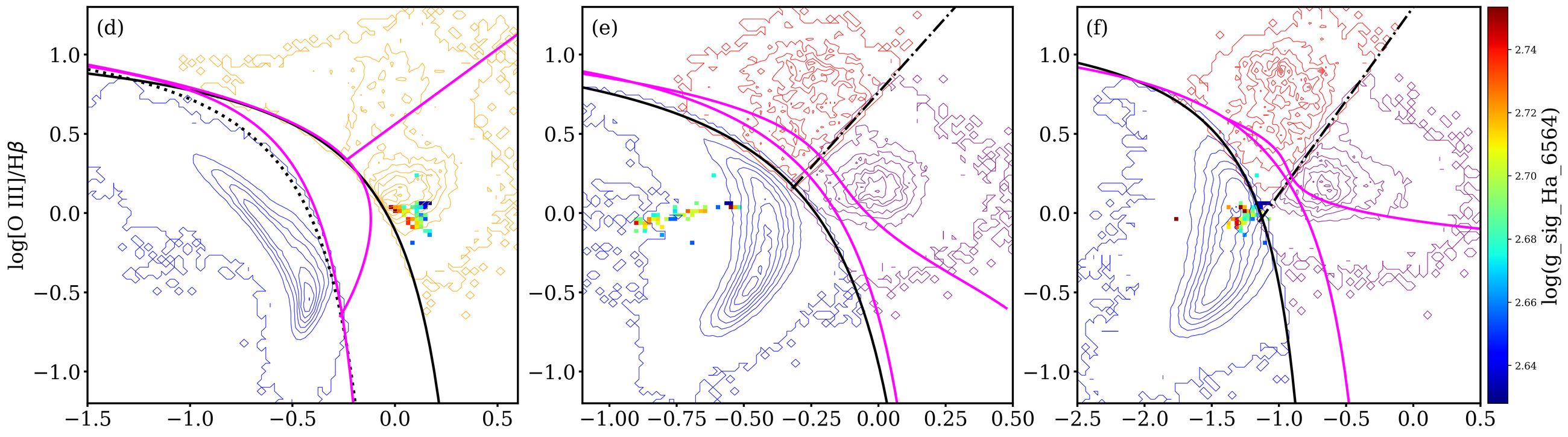}
\plotone{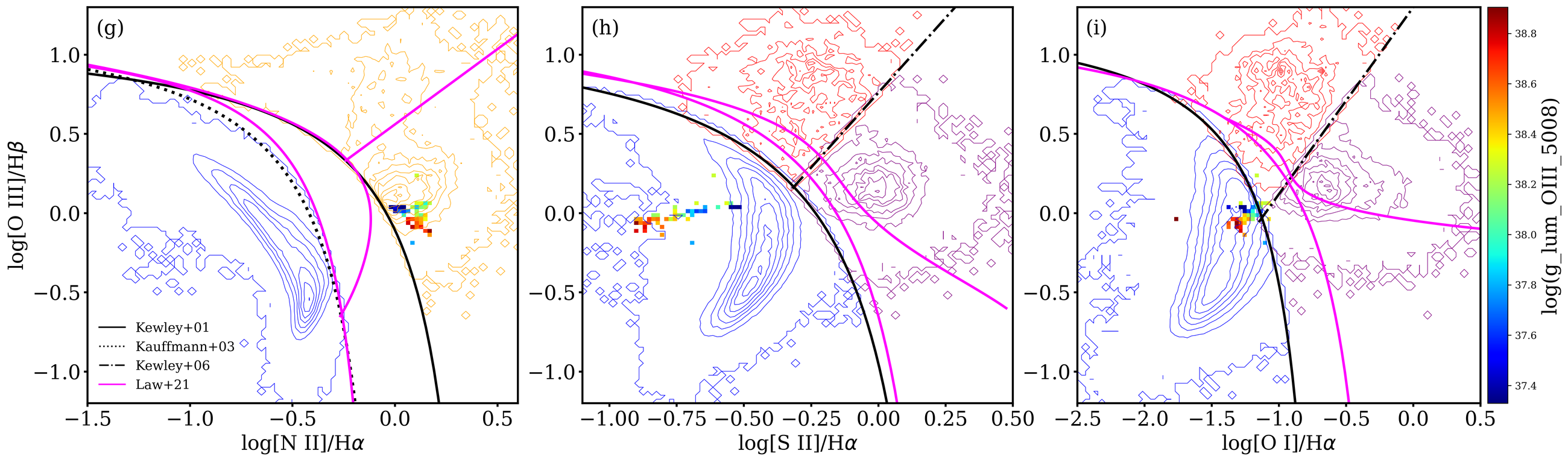}
\caption{Color-coded BPT diagrams showing the normalized elliptical radius} (top), H$\alpha$ velocity dispersion (middle), and [O III] luminosity (bottom) for the selected SF$^U$ spaxels. The uppermost spaxel on the [N II]- and [S II]-BPT diagrams, as well as the leftmost spaxel on the [O I]-BPT diagram, originates from a different galaxy, MaNGA ID \texttt{1-43214}, while the remaining 56 spaxels are from MaNGA ID \texttt{1-614567}.
\label{fig:umap_amb_bpt_sf_sub1_filt}
\end{figure*}

Figure \ref{fig:umap_amb_bpt_sf_sub1_filt} shows the 57 selected SF$^U$ spaxels that meet these criteria. As expected, the observed trends remain consistent: spaxels with broader H$\alpha$ emission lines are positioned deeper within the SF region of the [S II]-BPT diagram and are located farther from the galaxy center. Their location away from the center suggests that these spaxels are not strongly influenced by AGN activity. Additionally, their relatively low [O III] luminosity indicates that high star-formation activity alone may not fully explain the observations, suggesting other processes may be at play. A single extreme outlier—located at the upper envelope of the [N II]- and [S II]-BPT distributions and at the leftmost edge of the [O I]-BPT panel—stems from a different galaxy, MaNGA \texttt{1-43214}.

\begin{figure*}
\centering
\includegraphics[width=0.5\textwidth]{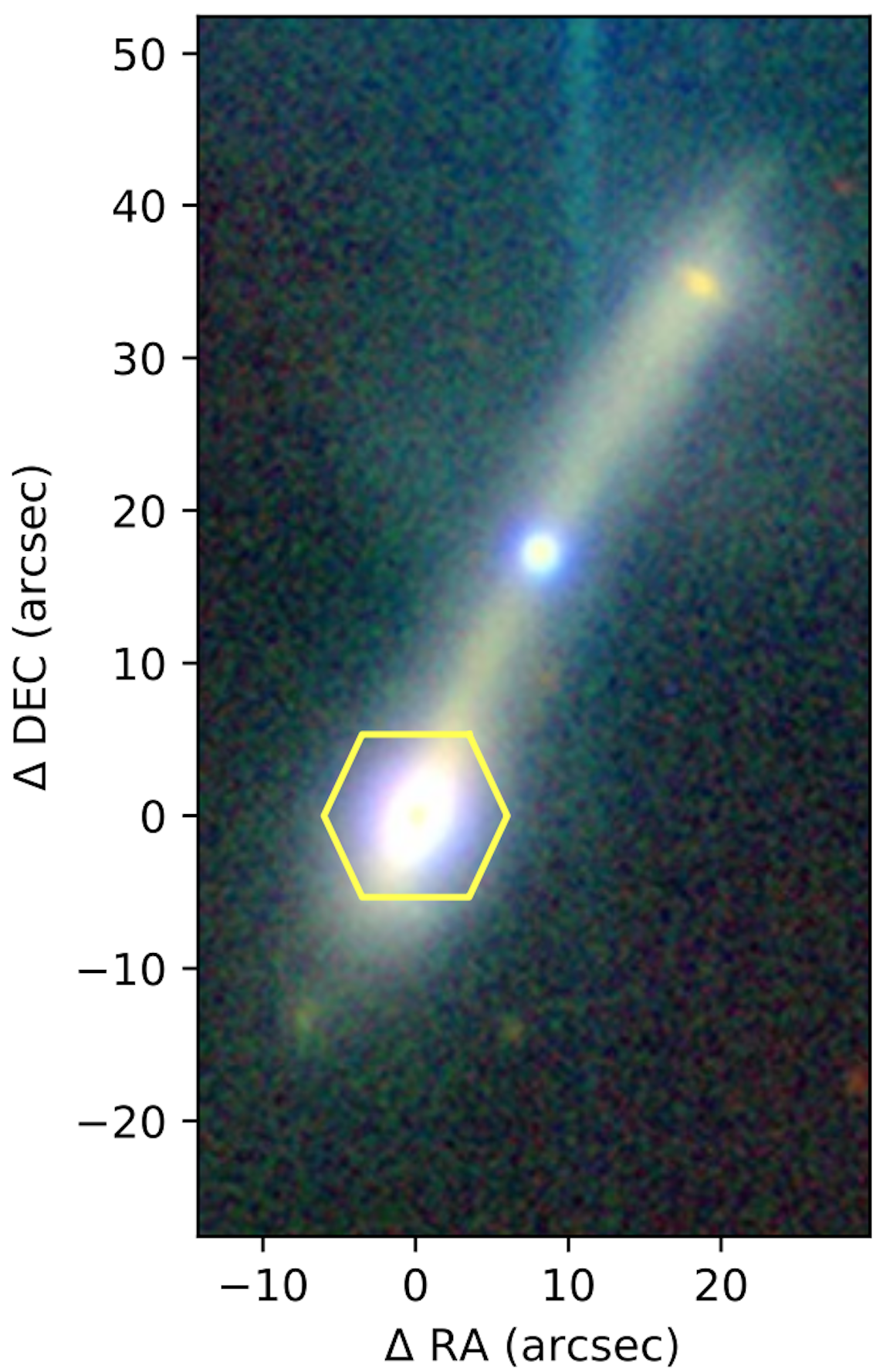}%
\caption{SDSS image of MaNGA \texttt{1-614567}. The yellow outline marks the MaNGA IFU field of view. The disturbed morphology is consistent with an ongoing interaction (and possibly a merger). This target contributes most of the selected SF$^U$ spaxels analyzed in Figure \ref{fig:umap_amb_bpt_sf_sub1_filt}.}
\label{fig:merging}
\end{figure*}

Figure \ref{fig:merging} shows the SDSS image of MaNGA \texttt{1-614567}, which exhibits clear signatures of interaction and a possible merger. Inspection of the MaNGA spectra for \texttt{1-614567} indicates a nuclear narrow-line Seyfert Type 1-like source with luminous broad Balmer components. The point spread function (PSF) can spread this nuclear light into adjacent spaxels, artificially broadening H$\alpha$ off-nucleus. This provides an alternative to shocks for the elevated H$\alpha$ in some SF$^U$ spaxels. In such cases, we expect Balmer lines to broaden preferentially, whereas shock broadening should affect both Balmer and forbidden lines; consequently, interpretations based solely on H$\alpha$ dispersion should be treated with caution unless supported by comparable broadening in, e.g., [O III] or [S II].

\bibliography{sample631}{}
\bibliographystyle{aasjournal}

\end{document}